\documentclass[11pt]{article}
\pdfoutput=1
\usepackage{jheparxiv}

\usepackage[dvipsnames]{xcolor}
\usepackage{tikz-feynman}
\usepackage{amsmath}
\usepackage{amsfonts}
\usepackage{bm}
\usepackage{braket}
\usepackage{comment}
\usepackage[normalem]{ulem}
\usepackage{amssymb} 
\usepackage{dsfont}
\usepackage{mathtools}
\usepackage{soul}
\usepackage{subcaption}

\newcommand{\ud}{{\mathrm{d}}}
\newcommand{\LCm}{{\scriptscriptstyle -}}
\newcommand{\LCp}{{\scriptscriptstyle +}}
\newcommand{\LCpm}{{\scriptscriptstyle \pm}}

\newcommand{\LCperp}{{\scriptscriptstyle \perp}}

\newcommand{\be}{\begin{equation}}
\newcommand{\ee}{\end{equation}}
\newcommand{\bi}{\begin{enumerate}}
\newcommand{\ei}{\end{enumerate}}
\newcommand{\ba}{\begin{aligned}}
\newcommand{\ea}{\end{aligned}}

\newcommand{\sfu}{{\mathsf u}}
\newcommand{\sfv}{{\mathsf v}}
\newcommand{\sfJ}{{\mathsf J}}
\newcommand{\sff}{{\mathsf f}}
\newcommand{\sfg}{{\mathsf g}}
\newcommand{\sfb}{{\mathsf b}}
\newcommand{\sfc}{{\mathsf c}}
\newcommand{\sfp}{{\mathsf p}}
\newcommand{\sfq}{{\mathsf q}}

\newcommand{\sfX}{{\mathsf x}}

\newcommand{\sfh}{{\mathsf h}}

\newcommand{\im}{{\mathrm{i}}}

\newcommand{\sft}{{\mathsf t}} 
\newcommand{\e}{{\mathrm e}}

\newcommand{\shock}{shock}
\newcommand{\shocks}{shocks}

\title{Radiation from scattering on multiple shockwaves}

\author[1]{Tim Adamo,}
\emailAdd{t.adamo@ed.ac.uk}

\author[2]{Anton Ilderton}
\emailAdd{anton.ilderton@ed.ac.uk}

\author[3]{\& Hannah Kingham}
\emailAdd{h.kingham@nbi.ku.dk}

\affiliation[1]{School of Mathematics and Maxwell Institute for Mathematical Sciences, \\
University of Edinburgh, EH9 3FD, United Kingdom}

\affiliation[2]{Higgs Centre for Theoretical Physics, School of Physics and Astronomy, \\
University of Edinburgh, EH9 3JZ, United Kingdom}

\affiliation[3]{Center of Gravity, Niels Bohr Institute, \\ Blegdamsvej 17, 2100 Copenhagen, Denmark}

\date{}
\abstract{It is well known that radiation produced by scattering an electron off multiple sources depends on the length of the mean free path between these scatterings. When this is sufficiently small, the spectrum is suppressed due to destructive interference, a famous result known as the Landau-Pomeranchuk-Migdal (LPM) effect. This is usually modelled approximately, with scattering centres treated linearly (e.g., as screened Coulomb potentials), high-energy probes and soft emitted photons. In this paper, we consider a scenario where multiple scatterings can be computed exactly: charged probe scattering on multiple null electromagnetic shockwaves. Here, the amplitudes for elastic scattering and single photon emission can be determined exactly, for arbitrary kinematics, in terms of the shock profiles in the transverse plane. We use these amplitudes to compute the radiation density (doubly-averaged over shock locations in the transverse plane and lightfront time) in the regimes of independent and interfering scatterings. For shock profiles which are radially symmetric in the transverse plane (including the ultraboosted Coulomb field), we recover the known LPM suppression, albeit with more general kinematics. In addition, our setup allows us to consider the source-free case of multiple scattering on impulsive/ultra-short plane waves, where we find much stronger suppression of the spectrum in the LPM regime.}

\begin{document}

\maketitle

\newpage

\section{Introduction}

Bremsstrahlung -- radiation induced by scattering -- is one of the most fundamental and well-studied phenomena in physics. It is particularly interesting to consider the problem of radiation arising from a probe particle undergoing multiple scatterings, as this is the underlying mechanism for radiation produced by scattering through an atomic or molecular medium (cf., \cite{Ter-Mikaelian:1972,Akhiezer:1996pa}). In the setting of quantum electrodynamics (QED), this problem was considered in the earliest days of the subject, over 90 years ago, by Bethe and Heitler~\cite{Bethe:1934za}. Assuming that the radiation produced from each scattering centre can be treated independently, they showed that the spectrum, or radiation density per unit length of the medium, grows logarithmically in a natural dimensionless parameter.

Twenty years later, Landau and Pomeranchuk~\cite{Landau:1953um,Landau:1953gr,Feinberg:1956} observed that the assumption of independent radiation from the multiple scattering centres is not universally valid. When the mean free path of the electron between successive scatterings is smaller than the wavelength of emitted photons, the emissions from distinct scattering centres will destructively interfere, leading to a suppression of the spectrum relative to the Bethe-Heitler result. Migdal subsequently developed a formalism which captured both the Bethe-Heitler regime of independent radiation and the suppression due to destructive interference as the medium becomes increasingly dense~\cite{Migdal:1956tc}. This suppression of radiation at high density in QED is now known as the \emph{Landau-Pomeranchuk-Migdal} (LPM) effect.

The LPM effect was directly observed over thirty years ago in the E-146 experiment at SLAC~\cite{Anthony:1995fs,SLAC-E-146:1997hnd}, and underpins important phenomena in the study of cosmic rays and neutrino detection (cf., \cite{Stanev:1982au,Zas:1991jv,Blumer:2009jrd,Gerhardt:2010bj,Anchordoqui:2018qom}). Furthermore, the generalisation of the LPM effect to quantum chromodynamics (QCD) (where the situation is qualitatively different, as emitted gluons continue to interact with the scattering centres) led to a resurgence of interest in the mid-1990s and the development of many improved derivations of the effect~\cite{Gyulassy:1993hr,Wang:1994fx,Perl:1994bg,Levin:1995mx,Baier:1996vi,Zakharov:1996fv,Zakharov:1998sv,Baier:1998kq}; see~\cite{Klein:1998du,Baier:2000mf} for contemporary overviews. All of these approaches incorporate some approximations (e.g., high-energy probes, soft photon emission), but the most ubiquitous is in the underlying model for multiple scattering. Each scattering centre is treated \emph{linearly} in the Born approximation, usually as a static screened Coulomb potential.   

\medskip

In this paper, we revisit the electromagnetic radiation induced by multiple scatterings through the framework of strong-field QED. The key feature of strong-field QED is that classical background electromagnetic fields are treated \emph{exactly and non-perturbatively}, with quantum electrons, positrons and photons described perturbatively around the fixed background through the Furry expansion~\cite{Furry:1951bef}. There are many experimental scenarios where a separation of scales necessitates the use of strong-field QED, from intense lasers to neutron star atmospheres (see~\cite{Gonoskov:2021hwf,Fedotov:2022ely} for reviews). However, experimental confirmation of the LPM effect -- based on the Born approximation of the background scattering centres -- makes it seem that this is not a scenario requiring strong-field treatment.

Nevertheless, there are several reasons why an exact description based in strong-field QED would be advantageous. Such a description would be more robust, holding for generic kinematics (of probe and emitted photons), encode all-order information beyond the Born approximation, and provide a concrete framework in which to incorporate corrections to the leading LPM effect (cf., ~\cite{Galitsky:1964,Arnold:2025dqj}). Furthermore, a strong-field setup can enable the treatment of radiation from multiple scatterings in new settings which are distinct from the screened Coulomb models typically used in modelling the LPM effect.

In the first instance, a strong-field model based on multiple Coulombic potentials is simply not practicable: even for a single centre, exact wavefunctions will look like confluent hypergeometric functions, making explicit scattering computations practically impossible.
This brings us to the canonical difficulty of strong-field QED, or indeed any background field theory: the challenge of finding a background field configuration which is a realistic model for the physics of interest while being simultaneously amenable to exact calculations.  

With the high-energy nature of the LPM effect in mind, we are motivated to consider a strong-field model of multiple scattering based on multiple null \emph{shockwaves}. These are electromagnetic fields supported on parallel lightfronts which are sourced by freely specified charge distributions in the transverse plane. Such shockwaves can be viewed as arising from ultraboosts of charged matter (e.g., the Coulomb field)~\cite{Bonnor:1969rb,Aichelburg:1970dh,Robinson:1984,Dray:1984ha,Ferrari:1988cc,Jackiw:1991ck}, but also include the source-free impulsive plane waves~\cite{Penrose:1972xrn}. The null nature of these backgrounds ensures that background-coupled wavefunctions for charged particles are WKB-exact~\cite{Penrose:1965rx,Jackiw:1991ck,Lodone:2009qe,Adamo:2021jxz}.

With these tools, we will compute both the elastic and single photon emission (i.e., ``non-linear Compton'') amplitudes at tree-level exactly in any multi-shockwave background. In both cases, the amplitudes are controlled by momentum transfer weighting functions, defined by the shockwave profiles in the transverse plane. Using these amplitudes, we compute the radiation density per unit length (in lightfront time). We evaluate this physical observable explicitly in the Bethe-Heitler and LPM regimes for two classes of multi-shock background: those which are radially symmetric in the transverse plane and the source-free impulsive plane waves. The former are the multi-shocks which mirror the standard setting of scattering through matter, as they include the ultraboosted Coulomb fields, while the latter can be viewed as a model for scattering through a train of ultra-short pulses of radiation.

For radially symmetric shocks, our strong-field QED setup reproduces the well-known results for both the Bethe-Heitler and LPM regimes, albeit allowing for more general kinematics and charge sources. While this can include all-order effects for generic sources, for shocks arising from ultraboosted Coulomb fields only the Born approximation contributes, giving exact matching with the standard LPM spectrum in this case~\cite{Baier:1996vi}. 

For the radiation produced by scattering on multiple impulsive plane waves, we observe a qualitative difference between the cases of a massive and massless probe particle. In both cases, the spectrum is suppressed relative to the Bethe-Heitler regime, with exponential suppression in the massive case and power-law suppression in the massless case. We also compare these findings with standard results in the strong-field QED literature on radiation from scattering in a constant crossed field, uncovering a remarkably precise correspondence with the LPM regime.

This paper is organised as follows. We give our conventions immediately below, then in Sec.~\ref{sec:shocks-and-amps} we introduce our multiple shock backgrounds and scattering setup,  calculate the elastic and photon emission amplitudes, and explore their properties. In Sec.~\ref{sec:observables} we turn to the computation of associated observables, including the emitted radiation density. This quantity is analysed for the cases of radially symmetric shocks in Sec.~\ref{sec:radial} and for impulsive waves in Sec.~\ref{sec:IPW}. We conclude in Sec.~\ref{sec:concs}.

\subsection*{Conventions and notation}
We use hats to represent factors of $2\pi$ in integration measures and delta functions, that is ${\hat \ud} p \equiv \ud p/(2\pi)$ and ${\hat \delta}(p) = 2\pi \delta(p)$. We work throughout in lightfont coordinates such that the Minkowski metric is
\be
    \ud s^2 = 2\,\ud x^\LCp \ud x^\LCm - \ud x^\LCperp \ud x^\LCperp= 2\,\ud x^\LCp \ud x^{\LCm}-|\ud x^{\LCperp}|^2\;,
\ee
where $x^\LCperp$ is a 2D (`transverse') vector and $x^\LCm$ is `lightfront time'. (In terms of Cartesian coordinates we have $x^\LCpm=(x^0 \pm x^3)/\sqrt{2}$ and  $x^\LCperp=\{x^1,x^2\}$.)

A vector $c_\LCperp$ in the transverse plane will frequently be denoted by the corresponding sans-serif symbol $\mathsf{c}$. The positive-definite inner product on the transverse plane can then be denoted as $|c_{\perp}|^2=\mathsf{c}^2$. It is sometimes notationally convenient to promote these to four-vectors $c_\mu$, understood as  $c_{\mu}\equiv\delta_{\mu}^{\perp}\, c_{\perp}$.

The on-shell integration measure for momentum $p_\mu$ in lightfront coordinates is written
\be
    \ud p_{\text{o.s.}} := \frac{\hat\ud p_\LCp}{2p_\LCp}\,\Theta(p_\LCp)\, \hat{\ud}^2p_\LCperp \;,
\ee
where $\Theta(p_+)$ is the Heaviside step function and the corresponding on-shell delta function is
\be
    \delta_{\text{o.s.}}(q-p) = 2q_\LCp\, {\hat\delta}_\LCp(q-p)\,{\hat\delta}^2(\mathsf{q}-\mathsf{p}) \;.
\ee

\section{Scattering on multiple shocks}\label{sec:shocks-and-amps}

Consider a highly boosted beam of charge travelling in the $z$-direction. In the ultrarelativistic limit, the collective Coulomb field of the beam may be approximated by~\cite{Bonnor:1969rb}
\be\label{potential-1-shock}
    A_\mu = - n_\mu\, \delta(x^\LCm)\, \Phi(x^\LCperp) \;,
\ee
in which $n_\mu$ is defined by $n\cdot x \equiv x^\LCm$, and $x^\LCm=0$ is the null hypersurface along which the beam (approximately) propagates, while $\Phi(x^\LCperp)$ encodes the matter distribution in the $x^\LCperp=(x,y)$ plane transverse to the propagation direction. The corresponding source is
\be
    J^\nu = \partial_\mu F^{\mu\nu} = n^\nu\, \delta(x^\LCm)\, \nabla_\LCperp^2\Phi(x^\LCperp) \;,
\ee
in particular, it is localized on the $x^{\LCm}=0$ lightfront and determined in the transverse plane by $\nabla_{\perp}^2\Phi(x^\LCperp)$.
This class of field includes the ultraboost of the Coulomb field of a point charge $Q$, for which
\be\label{Coulombshock}
\Phi^{\mathrm{Coulomb}}(x^{\perp})=\frac{Q}{4\pi}\,\log\!\left(\mu^2\,|x^\LCperp|^2\right)\,,
\ee
where $\mu$ is an arbitrary mass scale. Also included are the pure vacuum configurations for which $\nabla^2_{\perp}\Phi=0$, known as \emph{impulsive plane waves} (IPWs):
\be\label{IPW}
\Phi^{\mathrm{IPW}}(x^\LCperp)=a_{\perp}\,x^{\perp}\,,
\ee
for some constant 2-vector $a_{\perp}$. Impulsive plane waves can be viewed as the limit of generic plane wave solutions of Maxwell's equations corresponding to ultra-short wave duration~\cite{Fedotov:2013uja,Ilderton:2019vot}. From now on, we will refer to \emph{all} fields of the form \eqref{potential-1-shock}, for generic profile $\Phi(x^\LCperp)$, as (electromagnetic) \emph{shockwaves} or \emph{shocks}\footnote{This is somewhat abusive nomenclature, as, for instance, the impulsive plane wave is source-free and should not strictly be viewed as a shockwave.}.

\medskip

\begin{figure}
    \centering
 \includegraphics[width=0.4\linewidth,angle=-45]{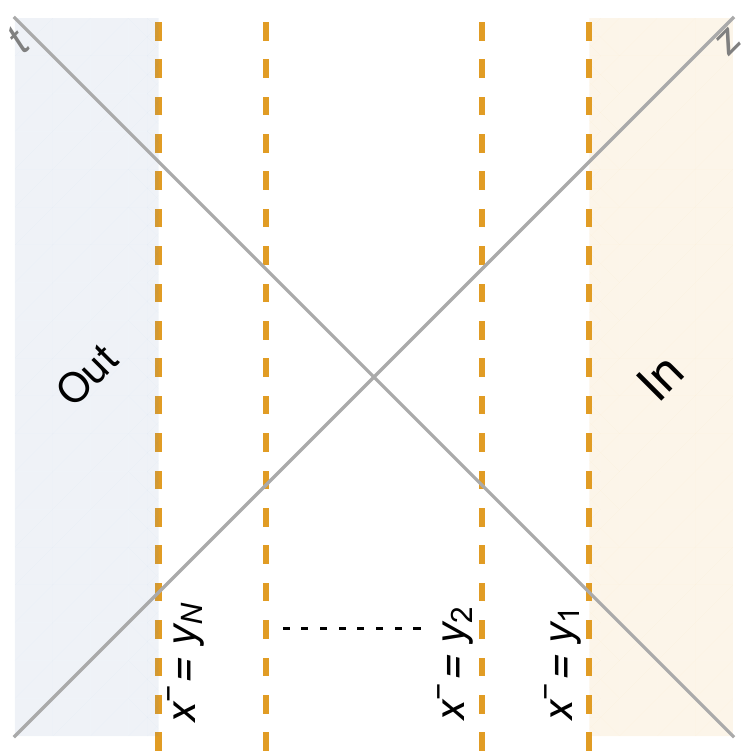}
    \caption{\label{fig:system} The fields of $N$ ultraboosted charge distributions are described by shockwave potentials of the form \eqref{potential-1-shock}, with support on the lightfronts $x^\LCm=y_1, \ldots, y_N$. We consider the scattering of massive and massless scalar particles on this background, with and without photon emission.}
\end{figure}

We consider the scattering of charged scalar probes (massive or massless) on $N$ \shocks\ with support at lightfront times $x^\LCm = y_j$, $j=1,2,\ldots, N$, and each with potentially different profiles $\Phi_j$. This multi-shock configuration is described by: 
\be\label{potential-N-shocks}
    A_\mu= - n_\mu\, \sum_{j=1}^N \delta(x^\LCm-y_j)\, \Phi_j(x^\LCperp) \;.
\ee
The setup is illustrated in Figure~\ref{fig:system}. 

\subsection{Wavefunctions}

The essential building blocks for calculations of scattering amplitudes in the multi-shock background are incoming and outgoing solutions $\varphi(x)$ of the Klein-Gordon equation
\be\label{KG}
    \big(D^2 + m^2\big)\varphi(x) = 0 \;, \qquad D_\mu = \partial_\mu + \im e\, A_\mu(x) \;,
\ee
in the background potential \eqref{potential-N-shocks}. Since the potential vanishes everywhere except on the \shocks, it is possible to construct solutions by patching superpositions of free solutions on either side of each \shock, with junction conditions fixed by continuity/cancellation of singularities in the Klein-Gordon equation, see e.g.~\cite{Penrose:1965rx,Lodone:2009qe,Adamo:2021jxz}. This procedure in the presence of a single shock (which we may freely place at $x^\LCm=0$) in the gauge \eqref{potential-1-shock} produces the wavefunction for an incoming particle of momentum $p_\mu$~\cite{Lodone:2009qe,Adamo:2021jxz}:
\be\label{solution-1-shock}
    \varphi_p(x) = \Theta(-x^\LCm)\, \e^{-\im\, p \cdot x} + \Theta(x^\LCm)\, \int \hat{\ud}^2\sfc\, W(\sfc)\, \e^{-\im\,\pi\cdot x},
\ee
in which $\Theta$ is the Heaviside step function, $W(\sfc)$ is the Fourier transform of the exponentiated potential,
\be\label{W-def}
    W(\sfc) := \int\!\ud^2\sfX\, \exp\!\left[-\im\, \sfc\cdot\sfX +\im\, e\, \Phi(\sfX)\right] \;,
\ee
and $\pi_\mu = [p-c]_\mu$, where the notation $[v]$ indicates the 4-vector with components $[v]_+=v_+$, $[v]_\LCperp=v_{\perp}$ and $[v]_{-}$ determined by the mass-shell condition. 
Since $\pi_\LCp=p_\LCp$, the probe lightfront momentum is conserved when crossing the \shock, while $\sfc$ parametrises the transverse momentum transfer to the probe. We discuss calculation of the `weight' $W(\sfc)$ in specific cases later, but note that the weight and wavefunction are manifestly all-orders in the coupling $e$. The interpretation of (\ref{solution-1-shock}) is that as a state of definite momentum $p$ crosses the \shock, it is kicked to a superposition of momentum eigenstates, with momentum $\pi_\mu\equiv\pi_\mu(p,\sfc)$: the probability amplitude to obtain a particular $\pi_\mu$ is simply $W(\sfc)$.

The generalisation to multiple shocks begins by first translating the single shock to an arbitrary lightfront time $x^\LCm = y_1$; substituting $x^\LCm \rightarrow x^\LCm - y_1$ into (\ref{solution-1-shock}) and dividing out by an overall phase to maintain the form $\e^{-\im p\cdot x}$ in the (asymptotic) past. The wavefunction becomes
\be\label{solution-shifted-shock}
    \varphi_p(x)\to \Theta(y_1-x^\LCm)\, \e^{-\im\, p \cdot x}
    +
    \Theta(x^\LCm-y_1) \int \hat{\ud}^2 \sfc_1\, W(\sfc_1)\, \e^{\im\,y_1(\pi^1 - p)_\LCm}\,
    \e^{-\im\,\pi^1 \cdot x}
    \;.
\ee
Here, we have relabelled $\sfc\to\sfc_1$ and $\pi\to \pi^1$; the only real difference compared to (\ref{solution-1-shock}) is the presence of an additional phase factor under the integral, given by the difference of the lightfront momenta of the probe before and after the \shock, multiplied by the shock position. This phase is, of course, obscured when a single \shock\ is placed at $x^\LCm=0$, but the \emph{accumulation} of such phases for multiple \shocks\ will be crucial later on.

So, given this continuation of a momentum eigenstate across a single \shock, we proceed to continue the resulting superposition of momentum eigenstates across subsequent \shocks\ following the same pattern. To write down the wavefunction in all of spacetime for an arbitrary number of shocks, we introduce the following notation, which will be of use throughout. Define $\pi^{k}_{\mu}$ to be the on-shell momenta
\be\label{pi-def1}
 \pi^k_\mu = [p-
 c_1-c_2-\ldots -c_k]_\mu \qquad
 \text{and}
 \qquad \pi^0_\mu = p_\mu \;,
\ee    
Then, starting from $\pi^0_\mu:=p_\mu$, the wavefunction of initial momentum $\pi_0$ is kicked to superpositions of $\pi^1_\mu$, then of $\pi^2_\mu$, and so on, as it crosses sequential shocks. When crossing the \shock\ at $y_j$ the wavefunction picks up the weight factor $W_j$ and the corresponding phase containing the difference of momenta. The final result is that the exact wavefunction is  
\be\label{incoming-solution}
   \varphi^{\text{in}}_p(x) =
   \begin{cases}
        \e^{-\im\,p\cdot x} \;, & x^\LCm < y_1 \;, \\[5pt]
        \displaystyle
        \left[\prod_{j=1}^k \int \hat{\ud}^2 \sfc_j\, W_j(\sfc_j)\, \e^{\im y_j\,(\pi^j_\LCm - \pi^{j-1}_\LCm)} \right]  \,\e^{-\im\,\pi^k \cdot x} \;,
        & y_k < x^\LCm < y_{k+1} \;, \\[15pt]
        \displaystyle
        \left[\prod_{j=1}^N \int \hat{\ud}^2 \sfc_j\, W_j(\sfc_j)\, \e^{\im y_j\,(\pi^j_\LCm - \pi^{j-1}_\LCm)} \right]\,  \e^{-\im\,\pi^N \cdot x} \;, & y_N < x^\LCm\;.
    \end{cases}
\ee
In particular, observe that before encountering the first shock at $x^-=y_1$, the wavefunction is simply a free momentum eigenstate of momentum $p_{\mu}$; as each successive shock is crossed the wavefunction becomes increasingly complicated combinations of plane waves.

The outgoing wavefunction for a particle with momentum $q$ in the asymptotic future has a similar structure: the wavefunction becomes a superposition of free states of acquired transverse momentum $+\sfc$ rather than $-\sfc$, and weighted by ${\overline W}$ rather than $W$ as it crosses each \shock, from the future to the past. The result is (taking a conjugate, as this is how the wavefunction will later appear in amplitudes) 
\be\label{outgoing-solution}
   {\bar\varphi}_q^{\text{out}}(x) =
   \begin{cases}
        \e^{\im\,q\cdot x} \;,
        & x^\LCm > y_N \;, \\[5pt]
       \displaystyle
       \left[\prod_{j=k+1}^{N}\int \!\hat{\ud}^2\sfc_j\, W_j(\sfc_j)\,
       \e^{-\im y^\LCm_{j}\,(\bar\pi^{N-j+1}_\LCm-\bar\pi^{N-j}_\LCm)} \right]\, \e^{\im\,\bar\pi^{N-k} \cdot x} \;,
       & y_{k+1} > x^\LCm > y_{k} \;, \\[15pt]
        \displaystyle 
         \left[\prod_{j=1}^{N}\int \!\hat{\ud}^2\sfc_j\, W_j(\sfc_j)\,
       \e^{-\im y^\LCm_{j}\,(\bar\pi^{N-j+1}_\LCm-\bar\pi^{N-j}_\LCm)} \right]\, \e^{\im\,\bar\pi^{N} \cdot x} \;,
       & y_1 > x^\LCm \;,
    \end{cases}
\ee
in which the ${\bar \pi}^k_\mu$ are, in analogy to (\ref{pi-def1}), the on-shell momenta
\be\label{pi-bar-def1}
{\bar\pi}^{N-k}_\mu = [q + c_{N}+c_{N-1}  + \ldots + c_{k+1}]_\mu \qquad \text{and} \qquad
{\bar\pi}^0_\mu = q_\mu \;.
\ee

\subsection{Elastic scattering amplitude}

Armed with these wavefunctions, we can consider a scalar particle scattering from momentum $p_\mu$ to momentum $q_\mu$ on the multi-shock background \eqref{potential-N-shocks} without any emission. The relevant `one-to-one' amplitude is
\be
    \mathcal{A}:=\bra{q}S\ket{p} \;,
\ee
where $S$ is the scalar QED S-matrix including the background. In terms of the wavefunctions introduced above, the tree-level amplitude may be expressed in terms of the Klein-Gordon inner product as
\be
    \mathcal{A} =
    \im\,\lim_{x^\LCm \rightarrow \infty}
    \int\!\ud x^\LCp\,\ud^2 x^\LCperp\,
    \e^{\im\, q\cdot x}\,
    \overset{\leftrightarrow}{\partial}_\LCp
    \varphi_p(x) \;,
\ee
although this can be equivalently calculated at any finite value of $x^\LCm$ in the `out' region of Fig.~\ref{fig:system}, i.e.~$x^\LCm>y_N$. 

Using \eqref{incoming-solution}, the calculation is direct:
\be\label{A2start}
\begin{split}
   \mathcal{A} &=
     \lim_{x^\LCm \rightarrow \infty}
    \int\!\ud x^\LCp\,\ud^2 x^\LCperp\,
    \left[\prod_{j=1}^N \int \hat{\ud}^2 \sfc_j\, W_j(\sfc_j)\, \e^{\im\, y_j\,(\pi^j_{-} - \pi^{j-1}_{-})} \right]\, \e^{\im\, q\cdot x}\, (p_\LCp+ q_\LCp)\,  \e^{-\im\,\pi^N \cdot x} \\
    &= 
    \left[\prod_{j=1}^N \int \hat{\ud}^2 \sfc_j \, W_j(\sfc_j)\,
    \e^{\im\,y_j\,(\pi^j_{-} - \pi^{j-1}_{-})} \right]\, \delta_{\text{o.s.}}(q - \pi^N) \;.
\end{split}
\ee
This expression tells us that lightfront momentum is conserved in elastic scattering (i.e., $p_+=q_+$), but transverse momentum is \emph{not}. Each $W_j(\sfc_j)$ acts as a probability amplitude for the probe to pick up transverse momentum $-\sfc_j$ at \shock\ $j$. 

While one set of $\sfc$-integrals can be performed against the on-shell delta function (as in the case of a single shock~\cite{Adamo:2021jxz}), it is more convenient to leave them unevaluated until considering probabilities and observables constructed from the modulus-square of the amplitude. The structure of this result will be revisited in Section~\ref{sec:shock-pw-comparison}.

\subsection{Photon emission amplitude}
We now consider scattering a scalar probe with incoming momentum $p_\mu$ to an outgoing scalar of momentum $q_\mu$ and an emitted photon of momentum $\ell_{\mu}$ and helicity $h=\pm1$. As the emitted photon does not couple to the multi-shock background (thanks to the linearity of Maxwell theory), its wavefunction is simply
\be\label{photonwf}
\varepsilon^{h}_{\mu}\,\e^{\im\,\ell\cdot x}\,,
\ee
where $\varepsilon^h_{\mu}$ is the on-shell polarization vector of helicity $h$. Of course, this requires a gauge fixing, which we take to be the Lorenz-lightfront gauge
\be\label{lorenz-lightfront}
\ell\cdot\varepsilon^h=0=n\cdot\varepsilon^h\,,
\ee
consistent with the symmetries of the multi-shock background.

With these conventions fixed, the $1\to2$ emission amplitude is given by:
\be\label{three-point-amp-start}
    \mathcal{M} := \bra{q,\ell^h}S\ket{p} = e \int\!\ud^4x\,
    {\bar\varphi}^{\,\text{out}}_{q} 
    \,
    \e^{\im\, \ell \cdot x}\, \varepsilon^{h\,\mu}\, \overset{\leftrightarrow}{\partial_\mu}\, 
    \varphi^{\text{in}}_{p}\,.
\ee
Plugging in the relevant incoming and outgoing wavefunctions and performing all integrals, one obtains
\be\label{basic-emission-amplitude}
\mathcal{M} = 
    -\im\,e
    \left[\prod_{j=1}^N \int\!\hat{\ud}^2\sfc_j\, W_j(\sfc_j)
    \right]\, 
     \delta_{\text{o.s.}}(q-\pi^N+\ell)\, \sum_{k=0}^N \frac{\varepsilon^h\cdot \pi^k}{\ell\cdot\pi^k}\,\e^{\im\, \sigma_k} \,\e^{\im\,\frac{\ell \cdot \pi^k}{p_\LCp-\ell_\LCp} y_a}\Big|_{a=k}^{a=k+1}\,,
\ee
where the momenta $\pi^k_{\mu}$ and $\bar{\pi}^k_{\mu}$, for $k=0,\ldots,N$ are defined in (\ref{pi-def1}) and (\ref{pi-bar-def1}). The phases
\be\label{phase-pattern}
    \sigma_k :=\sum_{j=k+1}^{N} y_j\, \big( {\bar\pi}^{N-j}_- -{\bar\pi}^{N-j+1}_{-} \big) 
    +
    \sum_{j=1}^k y_j\, \big(\pi^{j}_{-}-\pi^{j-1}_{-}\big) \;,
\ee
are analogous to the phases in the elastic case \eqref{A2start}, accumulated as the in and out wavefunctions cross \shocks. 

Finally, the difference of exponentials
\be\label{shockdifference}
\e^{\im\,\frac{\ell \cdot \pi^k}{p_\LCp-\ell_\LCp} y_a}\Big|_{a=k}^{a=k+1}:=\e^{\im\,\frac{\ell \cdot \pi^k}{p_\LCp-\ell_\LCp} y_{k+1}}-\e^{\im\,\frac{\ell \cdot \pi^k}{p_\LCp-\ell_\LCp} y_k}\,,
\ee
appearing in the emission amplitude (\ref{basic-emission-amplitude}) includes `fictitious' shock locations at $y_{0}=y_{N+1}:=\im\,\infty$, ensuring that contributions from the $x^{-}\to\pm\infty$ regions of integration are exponentially damped, consistent with the usual $\im\epsilon$-prescription. This notation also serves to keep the amplitude expression compact, as it allows contributions from the `in' and `out' regions to be conveniently written as part of the sum in \eqref{basic-emission-amplitude}. Note that in obtaining \eqref{basic-emission-amplitude} identities such as
\be\label{on-shellident}
\ell_{-}+\bar{\pi}^{N-k}_{-}-\pi^k_{-}=\frac{\ell\cdot\pi^k}{p_+-\ell_+}\,,
\ee
which hold on the support of the on-shell delta functions appearing under the integrals, have been used repeatedly.

\medskip

Although the kinematics of the probe scalar and emitted photon are completely arbitrary, the astute reader will have observed that the amplitude is nearly composed of sums over soft photon splitting functions associated with each shock crossing. Indeed, this is because everything non-trivial about the amplitude happens at the instants in lightfront time $x^-=y_1,\ldots,y_N$ when the probe crosses a shock. Investigating the soft limit of the emission amplitude sheds more light on this structure, and also serves as an easy consistency check.

For the purpose of studying the soft limit, let 
\be\label{soft1}
\ell_{\mu}=\omega\,\hat{\ell}_{\mu}=\frac{\omega}{\sqrt{2}}\left(1+|z|^2,\,z+\bar{z},\,-\im(z-\bar{z}),\,1-|z|^2\right)\,,
\ee
for $\omega$ the photon's frequency and $\hat{\ell}_{\mu}$ the null vector specified by a point $(z,\bar{z})\in S^2$ on the celestial sphere. The soft photon limit corresponds to $\omega\to 0$. Now, consider a generic term in the sum over over $k$ in \eqref{basic-emission-amplitude}, for $0<k<N$; in the soft limit this will go like
\be\label{soft2}
\frac{1}{\omega}\,\frac{\varepsilon^h\cdot\pi^k}{\hat{\ell}\cdot\pi^k}\,\e^{\im\,\sigma_k}\,\left(\e^{\im\,\omega\,\frac{\hat{\ell}\cdot\pi^k}{p_+}\,y_{k+1}}-\e^{\im\,\omega\,\frac{\hat{\ell}\cdot\pi^k}{p_+}\,y_{k}}\right)\,,
\ee
as $\omega\to 0$. Clearly, such terms are IR finite, as the $1/\omega$ pole is cancelled by the vanishing difference between the exponentials in the soft limit. Physically, this makes sense: these terms in the sum correspond to emission from an internal line in a Feynman diagram, which do not produce soft divergences at tree-level.

However, for the $k=0$ and $k=N$ terms in the sum, no such cancellation occurs, as there is only a single exponential term (the other corresponding to one of the fictitious shocks at infinity). This is the expected IR divergence associated with emission from an incoming or outgoing line. Therefore, we have
\be\label{soft3}
\lim_{\omega\to0}\mathcal{M}=\im\,e\,\left[\prod_{j=1}^{N}\int\hat{\ud}^2\sfc_j\,W_j(\sfc_j)\right] \delta_{\mathrm{o.s.}}(q-\pi^N) \left(\frac{\varepsilon^h\cdot\pi^N}{\ell\cdot\pi^N}\,\e^{\im\,\sigma_N}-\frac{\varepsilon^h\cdot\pi^0}{\ell\cdot\pi^0}\,\e^{\im\,\sigma_0}\right)+O(\omega^0)\,.
\ee
Now, using the delta functions, it follows that $\bar{\pi}^{N-j}_{-}=\pi^j_{-}$, which implies that $\sigma_0=\sigma_N$ in the soft limit. Using these facts along with the definition of $\sigma_N$ from \eqref{phase-pattern}, one finally obtains
\be\label{soft-limit}
\lim_{\omega\to0}\mathcal{M}=\im\,e\left(\frac{\varepsilon^{h}\cdot q}{\ell\cdot q}-\frac{\varepsilon^h\cdot p}{\ell\cdot p}\right)\mathcal{A}+O(\omega^0)\,,
\ee
where $\mathcal{A}$ is the elastic amplitude \eqref{A2start}. As desired, this is precisely the Weinberg soft photon theorem~\cite{Weinberg:1965nx}. While this is a useful consistency check on our emission amplitude, we stress that from now on we will consider photon emission with generic (non-soft) kinematics.

\subsection{Sourced vs.~source-free backgrounds}
\label{sec:shock-pw-comparison}

We end this section with a brief comment on the relations between amplitudes calculated in different shocks, i.e.~for different choices of the $\Phi_j(x^\LCperp)$. As observed before, the general class of multi-shocks \eqref{potential-N-shocks} considered here  contains both sourced and source-free fields. The latter case corresponds to impulsive plane waves (IPWs), with profiles given by $\Phi_j(x^\LCperp)=\mathsf{a}_j\cdot\mathsf{x}$, with $\mathsf{a}_j$ a constant 2-vector (as in \eqref{IPW}). This $\mathsf{a}_j$ is precisely the transverse momentum transferred to a classical particle crossing the $j^{\mathrm{th}}$ IPW -- see~\cite{Taub:1948zza,Sengupta:1949,Adamo:2017nia,Ilderton:2019vot,Adamo:2022rmp}.

There is a direct connection between amplitudes in the multi-IPW background and those in sourced multi-\shock\ backgrounds. This can be exposed by observing that the elastic scattering and photon emission amplitudes calculated above share some common structure: the same product of weight functions $W_j(\sfc_j)$ and integrals over the transverse momentum transfers $\sfc_j$ appears in both. Factorising these out, one obtains
\be
\begin{split}
    \mathcal{A} & =  \left[\prod_{j=1}^N \int\!\hat{\ud}^2\sfc_j\, W_j(\sfc_j)\right] \mathcal{A}_{\mathrm{IPW}}(\sfc_1,\ldots,\sfc_N)\,, \\
    \mathcal{M} & =  \left[\prod_{j=1}^N \int\!\hat{\ud}^2\sfc_j\, W_j(\sfc_j)\right] \mathcal{M}_{\mathrm{IPW}}(\sfc_1,\ldots,\sfc_N)    \,,
\end{split}
\ee
where $\mathcal{A}_{\mathrm{IPW}}(\sfc_1,\ldots,\sfc_N)$ and $ \mathcal{M}_{\mathrm{IPW}}(\sfc_1,\ldots,\sfc_N)$ are the elastic and emission amplitudes for a sequence of $N$ IPWs with transverse momentum transfers $\mathsf{a}_j=\sfc_j$. This is easily verified by direct computation of these amplitudes in a multi-IPW background, for which $W(\sfc)=\hat\delta_\LCperp(\sfc-\mathsf{a})$, as follows from \eqref{IPW} and (\ref{W-def}).

Thus, the elastic and photon emission amplitudes in sourced \shocks\ may be written as functional averages over the corresponding amplitudes in IPWs, which are source-free. The equivalent result for a single \shock\ was first observed in~\cite{Adamo:2021jxz}, and generalised to amplitudes for $N$-photon emission in~\cite{Copinger:2024twl}. For the extension of these `factorisation' results to loop level, see~\cite{Ilderton:2024ufp}.

\section{Observables}
\label{sec:observables}
%

At this stage we have the exact amplitudes for elastic \eqref{A2start} and single photon emission \eqref{basic-emission-amplitude} from a charged scalar probe scattering on the multi-shock background.
In this section, we turn to some basic observables which are computed from the elastic and photon emission amplitudes. Such observables include total and differential scattering probabilities, impulse/scattering angle, and the photon emission spectrum.

We will employ a certain averaging procedure motivated by the following physical concerns. Let us ask ourselves what the physical picture corresponding to the multi-shock setup would look like in a practical experimental setting. An obvious answer is that one is considering ultrarelativistic scattering of the probe through a medium approximated by the multi-shock configuration. In the source-free case, the picture is instead of a probe scattering through a wave region composed of multiple ultra-short pulses, modelled by IPWs. In these physical pictures, it is unlikely that one would have precise experimental control over the lightfront locations of the shocks or the transverse displacements of the probe as it passes through the medium/pulse configuration. Consequently, when computing observables we will average over the shock locations and transverse displacements; such averages are typical for the study of scattering through a medium (cf., \cite{Landau:1953gr,Landau:1953um,Migdal:1956tc,Akhiezer:1996pa,Baier:1996vi}).

After first reviewing the details of how amplitudes are dressed with wavepackets and averaged over transverse displacements, we compute the differential and total scattering probabilities in the elastic case, along with the impulse. In the case of photon emission, we focus on the radiated photon spectrum.

\subsection{Wavepackets and averages}

Let $\mathcal{M}=\mathcal{M}(p\to q+\cdots)$ denote a generic scattering amplitude for an incoming scalar probe of momentum $p$ to scatter to an outgoing probe of momentum $q$ plus any amount of radiation. Our elastic and single photon emission amplitudes are clearly special cases of such an amplitude. When computing observables, it is natural to supplement the amplitude by convolving the incoming state with a normalised wavepacket:
\be\label{wavepacket-amp}
    \mathcal{M}_\psi := \int\!\ud p_{\text{o.s.}}\, \psi(p)
    \, \e^{\im\sfb\cdot\sfp}
    \bra{q+\cdots}S\ket{p} 
    = \int\!\ud p_{\text{o.s.}}\, \psi(p) \, \e^{\im\, \sfb.\sfp}\,\mathcal{M} \;.
\ee
Here, $\sfb$ is a transverse impact parameter, while
$\psi(p)$ obeys the normalisation condition
\be
    \int\!\ud p_{\text{o.s.}}\, |\psi(p)|^2 = 1 \;.
\ee
Typically, one should think of $|\psi(p)|^2$ as a regulated on-shell delta function: that is, a normalised distribution strongly peaked around a certain chosen momentum. Furthermore, in observables which depend on the modulus-squared of the amplitude, integrating over the impact parameter $\mathsf{b}$ enables one to identify the wavepacket of the conjugate amplitude as the conjugate wavepacket of the unconjugated amplitude. 

\medskip

In the specific context of scattering on a multi-shock background, we will make the simplifying assumption that all of the shock profiles are \emph{identical}, up to displacements in the transverse plane:
\be\label{ident-shock}
\Phi_{j}(\mathsf{x})=\Phi(\mathsf{x}-\mathsf{b}_j)\,,
\ee
where $\Phi$ is the profile (now shared by all of the shocks) and $\mathsf{b}_j$ is the associated displacement in the transverse plane of the $j^{\mathrm{th}}$ shock relative to the origin. In Fourier space, this assumption corresponds to
\be
    W_j(\sfc_j) \to W(\sfc_j)\, \e^{-\im\,\sfc_j\cdot\sfb_j} \;,
\ee
at the level of the weights $W_j(\sfc_j)$ appearing in the amplitudes \eqref{A2start} and \eqref{basic-emission-amplitude}.

Of course, the choice of transverse displacements $\sfb_j$ for the shocks is entirely arbitrary, so they should be averaged\footnote{In the $N=1$ case, this averaging over $\sfb_1$ is equivalent to averaging over the probe impact parameter.} over with some probability density $\rho(\sfb_j)$. In particular, for any function (or distribution) $f=f(\sfb_1,\ldots,\sfb_N)$, define the average
\be\label{transverse-average-def}
    \langle f \rangle := \int\!\ud^2 \sfb_1\, \rho(\sfb_1) \cdots \int\!\ud^2 \sfb_N\, \rho(\sfb_N) \, f(\sfb_1,\ldots \sfb_N) \;,
\ee
with the normalisation of the distributions set by $\langle 1 \rangle = 1$. A typical assumption that one makes on the distribution $\rho$ is that it is broad in position space with respect to scales over which the probability differs significantly from zero. This means that one may approximate
\be\label{distribution}
\rho(\sfb_j)\simeq \frac{1}{V}\,,
\ee
for some large constant $V$, which can be thought of as the `volume of the transverse plane.'

Thus, the calculation of observables will typically involve integrating the probability density $|\mathcal{M}_{\psi}|^2$ over this flat distribution, which greatly simplifies the structure of the results. This is in complete analogy with how integrating over impact parameters simplifies probabilities and cross-sections in standard textbook discussions~\cite[\S 4]{Peskin:1995ev}. However, some care is required when manipulating these integrals, as one formally obtains factors of $0$ in the $V\to\infty$ limit which are balanced by infinite factors associated to the volume of the transverse phase space.

To see this more precisely, consider the following quantity which appears generically in the computation of probabilistic observables:
\begin{multline}\label{typical1}
\left\langle\prod_{j=1}^{N}\int\hat{\ud}^2\sfc_j\,\hat{\ud}^2\sfc'_{j}\,W(\sfc_j)\,\overline{W}(\sfc'_j)\,\e^{\im\,(\sfc'_j-\sfc_j)\cdot\sfb_j}\right\rangle \\
=\frac{1}{V^N}\,\prod_{j=1}^{N}\int\hat{\ud}^2\sfc_j\,\hat{\ud}^2\sfc'_{j}\,W(\sfc_j)\,\overline{W}(\sfc'_j)\,\hat{\delta}^2(\sfc'_j-\sfc_j) =\prod_{j=1}^{N}\frac{1}{V}\,\int\hat{\ud}^2\sfc_j\,|W(\sfc_j)|^2\,.
\end{multline}
In other words, after averaging over the $\sfb_j$,
\be\label{typical2}
\frac{1}{V}\,\int\hat{\ud}^2\sfc_j\,|W(\sfc_j)|^2\,,
\ee
is the basic \emph{distributional} quantity appearing in an observable.

Now, the weight $W(\sfc)$ defined by \eqref{W-def} is easily seen to obey the orthogonality relation
\be\label{W-orthogonality}
    \int\!{\hat \ud}^2\sfc\, W(\sfc)\,\overline{W}(\sfc-\sfc') = {\hat \delta}^2(\sfc')\;,
\ee
and so 
\be\label{W-norm}
    \int\!{\hat \ud}^2\sfc\, |W(\sfc)|^2 = {\hat \delta}^2(0)\;,
\ee
which is a divergent, ill-defined quantity. The standard resolution of this divergence is to balance it against the similarly divergent volume of the transverse plane:
\be\label{needed-norm}
        \frac{1}{V}\int\!\hat{\ud}^2\sfc\, |W(\sfc)|^2 = 1 \;.
\ee
A somewhat more flexible way of stating this is to replace the distributional combination
\be\label{reg-dist}
\frac{|W(\sfc)|^2}{V}\rightarrow|w(\sfc)|^2\,,
\ee
where 
\be\label{reg-dist2}
    \int\!{\hat \ud}^2\sfc\, |w(\sfc)|^2 = 1 \;, 
\ee
thus preserving the normalisation \eqref{needed-norm}. In particular, $|w(\sfc)|^2$ is a normalised probability density on transverse momentum transfers. 

The specific form of the regularised distribution $|w(\sfc)|^2$ obviously depends on the chosen shockwave profile $\Phi$. Later, we will see this explicitly when working with the specific examples of radially-symmetric shocks and impulsive plane waves.

\medskip

The final ingredient required in our computation of physical observables is an averaging over shock locations in lightfront time. In practice, one expects that the separation of the shocks will not be under precise control, but rather characterised by some typical separation scale, $\lambda$. This $\lambda$ is the mean free path of the probe scalar, which characterises the average lightfront time separating interactions between the probe and the background medium or pulse configuration.

Let $T_{j}:=y_{j+1}-y_j$ denote the lightfront separation of the background shocks in the exact amplitude computation. If $\langle f(y_1,\ldots,y_N)\rangle$ is some observable which has been averaged over transverse displacements according to \eqref{transverse-average-def}, then we define the `doubly-averaged' quantity
\be\label{lightfront-average}
\langle\!\langle f\rangle\!\rangle:=\prod_{j=1}^{N-1}\int_{0}^{\infty}\frac{\ud T_j}{\lambda}\,\e^{-T_j/\lambda}\,f(y_1,\ldots,y_N)\,.
\ee
In other words, we average over the separation of the shocks with an exponential distribution whose mean is $\lambda$.

\subsection{Elastic scattering}
Taking the elastic scattering amplitude $\mathcal{A}$ given by \eqref{A2start}, we convolve with a normalised wavepacket to form $\mathcal{A}_{\psi}$ according to \eqref{wavepacket-amp}. The differential elastic scattering probability is then simply
\be
    \frac{\ud \mathbb{P}_{1\to 1}}{\ud q_{\text o.s.}} = |\mathcal{A}_\psi|^2 \;.
\ee
More explicitly, one has
\begin{multline}\label{elastic-mess}
\frac{\ud \mathbb{P}_{1\to 1}}{\ud q_{\text o.s.}}=\int \ud p_{\text{o.s.}}\,\ud p'_{\text{o.s.}}\,\psi(p)\,\bar{\psi}(p')\,\e^{\im\,\sfb\cdot(\sfp-\sfp')} \,\prod_{j=1}^{N}\hat{\ud}^{2}\sfc_j\,\hat{\ud}^{2}\sfc'_j\,W(\sfc_j)\,\overline{W}(\sfc'_{j})\e^{\im\,\sfb_j\cdot(\sfc'_j-\sfc_j)} \\
\times\,\e^{\im\,y_j\,(\pi^{j}_{-}-\pi^{j-1}_{-})}\,\e^{-\im\,y_j\,(\pi^{\prime\,j}_{-}-\pi^{\prime\,j-1}_{-})}\,\delta_{\text{o.s.}}(q-\pi^{N})\,\delta_{\text{o.s.}}(q-\pi^{\prime\,N})\,,
\end{multline}
where $\pi^{\prime\,j}_{\mu}$ is defined by \eqref{pi-def1}, but with $p_{\mu}\to p'_{\mu}$ and $\sfc_j\to\sfc'_j$ in the obvious way.

At this point, one could proceed by using the on-shell delta functions to fix all of the longitudinal (plus) components of the momenta in the wavepackets, and one set of perpendicular momenta, say $\sfp'$. However, the integrals over $\sfp$ remain, along with the wavepackets. Instead, we average over the transverse displacements of the individual shocks according to \eqref{transverse-average-def} -- aside from the physical motivation for doing so, discussed above, this will allow us to remove the wavepackets, as follows. Averaging results in:
\be\label{el-diff-cross1}
\begin{split}
 \frac{\ud \langle \mathbb{P}_{1\to 1}\rangle}{\ud q_{\text o.s.}} 
 &=
 \int\prod_{i=1}^{N} \ud^{2}\sfb_j\,\rho(\sfb_j)\, |\mathcal{A}_\psi|^2 \\
 &= \int\ud p_{\text{o.s.}}\,|\psi(p)|^2\,\prod_{j=1}^{N}\hat{\ud}^{2}\sfc_j\,|w(\sfc_j)|^2\,\delta_{\text{o.s.}}(q-\pi^N)\,,
\end{split}
\ee
where the second line follows after performing the $\sfb_j$ integrals to generate a set of delta functions which, when combined with $\delta_{\text{o.s.}}(q-\pi^{\prime\,N})$, fix $p'_{\mu}=p_{\mu}$ and $\sfc_j=\sfc'_j$, allowing the primed variables to be integrated out. The normalised distribution $|w(\sfc)|^2$ is defined by regulating $|W(\sfc)|^2/V$ as in \eqref{W-orthogonality} -- \eqref{reg-dist2}. 

At this point, we can make the standard assumption that $|\psi(p)|^2$ is sharply peaked around a particular momentum, which we will (in another small abuse of notation) denote $p_{\mu}$. This allows us to factorise and discard the wavepacket contribution, as desired, leaving
\be\label{el-diff-cross2}
\frac{\ud \langle \mathbb{P}_{1\to 1}\rangle}{\ud q_{\text o.s.}}
=\prod_{j=1}^{N}\int\hat{\ud}^{2}\sfc_j\,|w(\sfc_j)|^2\,\delta_{\text{o.s.}}(q-\pi^N)\,.
\ee
Note that all of the phases present in the amplitude \eqref{A2start} drop out of this result because of the averaging, hence any information on the lightfront time locations $y_j$ of the \shocks\ is lost. As such there is no need to average over the shock locations in the elastic setting -- though this will certainly not be the case when emission is included, as we will see later.

\medskip

The total elastic scattering probability is obtained by integrating over $q$, which is made trivial by the remaining on-shell delta function in \eqref{el-diff-cross2}, giving
\be\label{elastic-amp-integrated1}
\begin{split}
    \langle\mathbb{P}_{1\to 1}\rangle
    &= \prod_{j=1}^N \int\! \hat{\ud}^2 \sfc_j\, |w(\sfc_j)|^2 \;.
\end{split}
\ee
Using the normalisation condition \eqref{reg-dist2} on the distribution $|w(\sfc)|^2$, it follows that 
\be\label{elastic-hmmmm2}
 \langle\mathbb{P}_{1\to 1}\rangle = 1 \;.
\ee
All parameter information has washed out of the \textit{total} probability, which is simply unity. While this may seem surprising at first, it is actually an expected feature: at this order in the background field expansion (i.e., $O(e^0)$), elastic scattering is the \emph{only} process which can occur. Thus, unitarity dictates that the total probability is equal to unity. In fact, using \eqref{W-orthogonality} is is possible to show the total probability is unity even without averaging over impact parameters (as it must be).

\medskip

The role of $|w(\sfc)|^2$ as a normalised probability density on the transverse momentum transfers is most clearly illustrated by considering the (averaged) \emph{impulse}: the expectation value of the change in momentum of a probe crossing the multi-shock background. This is computed from the (averaged) differential probability as
\be\label{impulse}
\begin{split}
    \Delta \sfp
    &=
    \int\!\ud q_{\text{o.s.}}\, (\sfq-\sfp)\,\frac{\ud \langle\mathbb{P}_{1\to 1}\rangle}{\ud q_{\text o.s.}} \\
    &= -\left[\prod_{j=1}^N \int\!  \hat{\ud}^2 \sfc_j\, |w(\sfc_j)|^2 \right] \sum_{k=1}^N \sfc_k \\
&    =
    - N\int\!  \hat{\ud}^2 \sfc\, |w(\sfc)|^2\, \sfc\, \;.
\end{split}
\ee
In the second line, the normalisation condition \eqref{reg-dist2} allows us to evaluate all but one integral in each term of the sum, and since each \shock\ has the same averaging, the final impulse is simply $N$ times that obtained from crossing a single \shock. As expected, $|w(\sfc)|^2$ serves as the probability distribution for the amount of momentum transfer imparted from each of the identical shocks.

\medskip

Rather than go through examples of the impulse calculated in different backgrounds (see e.g.~\cite{Adamo:2021jxz,Adamo:2022rmp}), let us make a quick but more general comparison with the `standard' computation of observables in the framework of ultrarelativistic scattering through a medium~\cite{Migdal:1956tc,Ter-Mikaelian:1972,Berestetskii:1982qgu,Gyulassy:1993hr,Wang:1994fx,Akhiezer:1996pa,Baier:1996vi}. The key distinction between our calculation and those which have appeared many times in the literature is that, here, observables are based upon an \emph{exact} amplitude computation (i.e., the elastic scattering amplitude in the multi-shock background), whereas the standard approach is based on an approximate scattering amplitude, where each centre is treated linearly as a potential.

We will see further distinction between the exact-versus-approximate treatment of the scattering centres in the presence of emitted radiation, but in the elastic regime the key distinction is that in the `standard' approximation, the centres are described by static potentials. There is no time-ordering of the centres, and scattering can occur in any ordering of their longitudinal separations. Contributions from orderings other than the identity are due to fluctuations of the vacuum into electron-positron pairs, but in the high-energy/ultrarelativistic limit, these pair production processes are suppressed, and only the unitary elastic process contributes (cf., \cite{Baier:1996vi}). In our calculation, the ultrarelativistic limit is already built into the background, as the sources are \emph{null} -- that is, they are `static' with respect to lightfront time $x^-$, rather than $t$. Such null backgrounds cannot pair-produce, and it is clear from the form of the probe wavefunctions, e.g.~(\ref{incoming-solution}), that the shocks are encountered `in order' by the probe. This explains why $\langle\mathbb{P}_{1\to1}\rangle=1$ exactly. This can also be seen as arising from the vanishing of `Z-diagrams' in lightfront perturbation theory~\cite{Brodsky:1997de,Heinzl:2000ht}.

\subsection{Photon emission}
%
In the case of photon emission, the obvious radiative observable is the photon spectrum, or equivalently, the differential probability of photon emission. This is given by
\be\label{def:diff-emission}
    \frac{\ud {\mathbb P}_{1\to2}}{\ud \ell_{\text{o.s.}}} = \sum_{h}\int\!\ud q_{\text{o.s.}}\, |\mathcal{M}_\psi|^2 \;,
\ee
where the sum in $h$ is over the emitted photon's helicity and $\mathcal{M}_{\psi}$ is the convolution of the emission amplitude \eqref{basic-emission-amplitude} against a wavepacket for the incoming electron, as defined by \eqref{wavepacket-amp}. As in the elastic case, the computation is made significantly easier by averaging over impact parameter of the incoming particle and transverse locations of the shocks; this allows the normalised wavepackets to be stripped off leaving
\begin{multline}\label{avg-diff-emission}
\frac{\ud \langle\mathbb P_{1\to2}\rangle}{\ud \ell_{\text{o.s.}}}=e^2\sum_{h}\int\ud q_{\text{o.s.}}\,\prod_{j=1}^{N}\hat{\ud}^2\sfc_j\,|w(\sfc_j)|^2\,\delta_{\text{o.s.}}(q+\ell-\pi^N)\,(p_+-\ell_+) \\
\times \,\left|\sum_{k=0}^{N}\frac{\varepsilon^h\cdot\pi^k}{\ell\cdot\pi^k}\,\e^{\im\,\sigma_k}\left(\e^{\im\,\frac{\ell\cdot\pi^k}{p_+-\ell_+}\,y_{k+1}}-\e^{\im\,\frac{\ell\cdot\pi^k}{p_+-\ell_+}\,y_{k}}\right)\right|^2\,,
\end{multline}
for the differential probability.

At this point, it is convenient to introduce photon momentum variables which are naturally adapted to the null geometry of our system. These are
\be\label{photon-vars}
s:=\frac{\ell_+}{p_+}\,, \qquad \sfu:=\frac{1}{\ell_+}\,(\ell_1,\,\ell_2)\,,
\ee
which are, respectively, the fraction of the initial electron lightfront energy carried away by the photon and a vector encoding the photon scattering angles\footnote{To see this we can set $p_\LCp=0$ for simplicity (corresponding to a `head-on' collision between the probe and the \shocks) and write $\sfu$ in Cartesian coordinates as $\sfu = \frac{\sqrt{2}\sin\theta}{1+\cos\theta}(\cos\phi,\sin\phi)$, where $\phi$ is the azimuthal angle in the transverse plane and $\theta$ is the polar angle relative to the \shock\ propagation direction.}.
Lightfront momentum conservation restricts $s$ to the range $0<s<1$, which we will keep implicit from now on.

Written in terms of these variables, the integral over the on-shell phase space of the outgoing electron can be performed against the on-shell delta functions. Performing the sum over photon helicities, the averaged photon spectrum is then given by:
\be\label{avg-spec1}
\frac{\ud \langle\mathbb P_{1\to2}\rangle}{\ud \ell_{\text{o.s.}}}=
4
e^2\,\frac{(1-s)}{\ell_+^2}\,\Theta(p_+-\ell_+)\,\left[\prod_{j=1}^{N}\int\hat{\ud}^2\sfc_j\,|w(\sfc_j)|^2\right]\left|\sum_{n=1}^{N}\mathsf{A}_{n}\,\e^{\im\,\varphi_{n}}\right|^{2}\,,
\ee
where
\be\label{Adef}
\mathsf{A}_{n}:=\frac{\sfu_{n}}{\sfu_n^2+\frac{m^2}{p_+^2}}-\frac{\sfu_{n-1}}{\sfu_{n-1}^2+\frac{m^2}{p_+^2}}\,,
\ee
for the shifted transverse momentum variables
\be\label{un-def}
 \sfu_n := \sfu
    - \frac{(\sfp-\sfc_1-\ldots -\sfc_n)}{p_\LCp} \;,
\ee
and phases
\be\label{phidef}
\varphi_n:=-\frac{\ell_+}{2\,(1-s)}\,\sum_{k=n}^{N-1}(y_{k+1}-y_k)\left(\sfu_k^2+\frac{m^2}{p_+^2}\right)\,.
\ee
Here the completeness relation has been used to perform the sum over photon helicities, and the modulus-square of $\sum_n\mathsf{A}_n \e^{\im\varphi_n}$ is understood to be the mod-square dot product in transverse space.

At this point, we focus our attention to the partially integrated spectrum
\be
    s\, \frac{\ud\langle\mathbb{P}_{1\to 2}\rangle}{\ud s}=\int\frac{\ud^2\ell_\LCperp}{4\pi^3}\,\frac{\ud \langle\mathbb P_{1\to2}\rangle}{\ud \ell_{\text{o.s.}}} \;.
\ee
This is essentially the spectrum in lightfront momentum $\ell_\LCp = s p_\LCp$ rather than in $\ell_0$, but multiplying through by $s$ as we have done here gives the natural equivalent of the intensity distribution for the null multi-shock geometry. The resulting quantity is:
\be\label{avg-spec2}
s\, \frac{\ud\langle\mathbb{P}_{1\to 2}\rangle}{\ud s}=\frac{\alpha}{\pi^2}\,(1-s)\,\int\ud^2\sfu\,\left[\prod_{j=1}^{N}\int\hat{\ud}^2\sfc_j\,|w(\sfc_j)|^2\right]\sum_{n,m=1}^{N}\mathsf{A}_{n}\cdot\mathsf{A}_{m}\,\e^{\im\,\varphi_{nm}}\,,
\ee
where $\alpha$ is the fine structure constant and the phases are:
\be\label{phi_nm}
\varphi_{nm}:=\frac{\ell_+}{2\,(1-s)}\,\sum_{k=n}^{m-1}(y_{k+1}-y_k)\left(\sfu_k^2+\frac{m^2}{p^2_+}\right)\,,
\ee
when $m-1>n$. If $n>m-1$, the phase is identical up to an overall minus sign and swapping the limits of summation. In order to proceed further, we must pass to the regime of a large number of shocks, which significantly simplifies the sums still appearing in the spectrum\footnote{This calculation closely follows that in~\cite{Baier:1996vi}.}.

We first separate out the incoherent piece of the sum, writing 
\be\label{incoh-sum}
\sum_{n,m=1}^{N}\mathsf{A}_{n}\cdot\mathsf{A}_{m}\,\e^{\im\,\varphi_{nm}}=\sum_{\substack{m,n=1 \\ n\neq m}}^{N}\mathsf{A}_{n}\cdot\mathsf{A}_{m}\left(\e^{\im\,\varphi_{nm}}-1\right)+\left(\sum_{n=1}^{N}\mathsf{A}_n\right)^2\,,
\ee
and look for the leading behaviour of these sums in the $N\to\infty$ limit. Now, the definition \eqref{un-def} gives
\be
\sum_{n=1}^{N}\mathsf{A}_n=\frac{\sfu_N}{\sfu_{N}^2+\frac{m^2}{p_+^2}}-\frac{\sfu}{\sfu^2+\frac{m^2}{p^2_+}}\,,
\ee
so that the second sum on the right-hand-side of \eqref{incoh-sum} depends only on the combinations $\sfu$ and $\sfu_N$. Under the integrals in \eqref{avg-spec2}, one can then change variables to eliminate all $N$-dependence from this sum. Using the reality of the $\mathsf{A}_n$ and the definition of the phases \eqref{phi_nm}, it also follows that the first sum on the right-hand-side of \eqref{incoh-sum} can be written
\be
\sum_{\substack{m,n=1 \\ n\neq m}}^{N}\mathsf{A}_{n}\cdot\mathsf{A}_{m}\left(\e^{\im\,\varphi_{nm}}-1\right)=2\,\Re \sum_{n=1}^{N}\sum_{m=n+1}^{N}\mathsf{A}_{n}\cdot\mathsf{A}_{m}\left(\e^{\im\,\varphi_{nm}}-1\right)\,.
\ee
In this expression, consider the second summation over $m$ for some fixed $n$. The symmetries of the $\mathsf{A}_n$, and the ability to change variables in the $\sfc_j$ integrals means that, in the large $N$ limit, this sum returns $N$ copies of the result for $n=1$. 

Hence, we have that
\be\label{largeN}
\lim_{N\to\infty}\sum_{n,m=1}^{N}\mathsf{A}_{n}\cdot\mathsf{A}_{m}\,\e^{\im\,\varphi_{nm}}=2N\,\Re \sum_{n=0}^{\infty}\mathsf{A}_{1}\cdot\mathsf{A}_{n+2}\left(\e^{\im\,\varphi_{1(n+2)}}-1\right)+O\!\left((N^{-1})^0\right)\,.
\ee
The leading factor of $N$ can be accounted for by normalising the observable by the total length of the medium/shock train. More specifically, we can consider the differential radiation density in units of length associated to the mean free path of the probe electron between shocks, $\lambda$. This is given by
\begin{multline}\label{rad-dens1}
s\,\frac{\ud^2\langle\mathbb{P}_{1\to2}\rangle}{\ud x^{-}\,\ud s}=
\lim_{N\to\infty}\frac{s}{\lambda\,N}\,\frac{\ud\langle\mathbb{P}_{1\to2}\rangle}{\ud s} \\
=\frac{2\,\alpha}{\pi^2\,\lambda}\,(1-s)\int\ud^2\sfu\,\Re\,\sum_{n=0}^{\infty} \left[\prod_{j=1}^{n+2}\int\hat{\ud}^2\sfc_j\,|w(\sfc_j)|^2\right]\mathsf{A}_1\cdot\mathsf{A}_{n+2}\left(\e^{\im\,\varphi_{1(n+2)}}-1\right)\,,
\end{multline}
making use of \eqref{largeN}. 

Finally, we perform the averaging \eqref{lightfront-average} over the separation of the shocks; this integration acts only on the phases $\varphi_{1(n+2)}$. The result is the doubly-averaged radiation density:
\be\label{rad-dens2}
s\,\frac{\ud^2\langle\!\langle\mathbb{P}_{1\to2}\rangle\!\rangle}{\ud x^{-}\,\ud s}=\frac{2\,\alpha}{\pi^2\,\lambda}\,(1-s)\,\int\ud^2\sfu\,\Re\,\sum_{n=0}^{\infty} \left[\prod_{j=1}^{n+2}\int\hat{\ud}^2\sfc_j\,|w(\sfc_j)|^2\right]\mathsf{A}_1\cdot\mathsf{A}_{n+2}\left(\left[\prod_{k=1}^{n+1}\frac{1}{1-\im\,\theta_k}\right]-1\right)\,,
\ee
where
\be\label{thetadef}
\theta_{k}:=\frac{\lambda}{2}\,\frac{\ell_+}{(1-s)}\left(\sfu_k^2+\frac{m^2}{p_+^2}\right)\,.
\ee
The precise results for this observable -- as we will see -- depend upon the shock profiles, kinematics and mean free path, but there is already an important general distinction from the elastic observables which is clear. In the elastic case, all dependence on phases from the amplitude \eqref{A2start} dropped out of the observables. However, in the radiative case, the phases appearing in the photon emission amplitude \eqref{basic-emission-amplitude} are not lost in the photon spectrum: they feed through into the quantities $\theta_k$ appearing in \eqref{rad-dens2}. These quantities will be crucial in what follows. Tracing their origin, $\theta_k$ arises through the difference of particle energies (multiplied by the average shock separation $\lambda$) before and after photon emission between the $k^\text{th}$ and $(k+1)^\text{th}$ shocks.  That is, $\theta_k$ may equivalently be written as
\be
    \theta_k = \lambda\left( \ell_\LCm + [\pi^k-\ell]_\LCm - \pi^k_\LCm\right) =   \lambda \,\frac{\ell\cdot\pi_k}{p_\LCp-\ell_\LCp} \;.
\ee
These are precisely the sort of `accumulation phases' contributing to path interference effects for processes in background fields, see for example~\cite{Hebenstreit:2009km,Akkermans:2011yn,Ilderton:2019ceq}. It is also worth emphasising that the kinematics of the incoming electron and emitted photon are totally \emph{arbitrary}: this is one payoff for basing the computation on an exact amplitude in background field theory.

\section{Radially symmetric shocks}
\label{sec:radial}
We now turn to evaluation of the doubly-averaged radiation density \eqref{rad-dens2} for specific classes of multi-shock backgrounds. For our first example, we consider massless probe scattering ($m=0$) and shock profiles which are radially symmetric in the transverse plane: $\Phi(\sfX)=\Phi(|\sfX|)$. Such profiles are the natural ultra-boosted versions of the (screened) Coulomb charges typically used in the literature to model scattering through a medium~\cite{Migdal:1956tc,Akhiezer:1996pa,Gyulassy:1993hr,Wang:1994fx,Baier:1996vi}; the basic example is the ultraboost of the point charge \eqref{Coulombshock}. Although our calculation is based on the \emph{exact} photon emission amplitude in the multi-shock background, we will see that eventually the computation can be essentially pattern-matched onto the well-known results in the literature, due to the shared symmetries.

After identifying relevant parameters and rewriting the general formula for the radiation density in a useful way, we proceed to evaluate the spectrum in two regimes, which are referred to as the \emph{Bethe-Heitler} (BH)~\cite{Bethe:1934za} and \emph{Landau-Pomeranchuk-Migdal} (LPM)~\cite{Landau:1953gr,Landau:1953um,Migdal:1956tc} regimes. These correspond to the regimes where the multishock background and kinematics are such that radiation is produced independently at each shock (BH), or there is destructive interference between radiation from different shocks (LPM).

\subsection{Identifying parameters and adapting the spectrum}
%
With a radially symmetric shock profile, it follows that the transverse momentum distribution is also radially symmetric, so that $|w(\sfc)|^2 \to |w(\sfc^2)|^2$. Assuming that this distribution is peaked around some scale $|\sfc|=\mu$, one can work with dimensionless momentum transfer variables $\sft=\sfc/\mu$. The transverse photon variables can similarly be rescaled by setting $\sfu=(\mu/p_+)\,\sfv$, where now $\sfv$ still encodes the photon scattering angles, albeit with a different overall scaling. Assuming (without loss of generality) that we work in a frame wherein the incoming electron has zero transverse momentum ($\sfp=0$), it follows that in the massless limit
\be\label{v-perp1}
\sfu_k=\frac{\mu}{p_+}\left(\sfv +\sft_1+\cdots+\sft_k\right):=\frac{\mu}{p_+}\,\sfv_k\,.
\ee
In these variables, the phases \eqref{thetadef} become
\be\label{theta-k-radial}
    \theta_k = \frac{\lambda\, \mu^2}{2\, p_\LCp}\,
     \frac{s}{1-s}\,\sfv_k^2 \equiv \kappa\,\sfv_k^2\,.  
\ee
The dimensionless quantity
\be\label{kappa-radsym}
\kappa=\frac{\lambda\, \mu^2}{2\, p_\LCp}\,
     \frac{s}{1-s}\,,
\ee
is the only combination of parameters and photon momentum variables which will enter non-trivially in the final result for the radiation density.

It is worth commenting briefly on the physical interpretation of $\kappa$. For fixed photon and probe kinematics, $\kappa\sim\lambda$, the mean free path of the probe electron between shocks. So one can intuitively think of $\kappa$ as the quantity which controls `average separation' of the radially-symmetric shocks in lightfront time. One therefore expects that $\kappa$ will control the regimes of emitted radiation, with the independent BH regime corresponding to large $\kappa\gg1$ and the interference LPM regime corresponding to small $\kappa\ll1$.

\medskip

At this point, it is useful to adapt the expression for the radiation density \eqref{rad-dens2} to the setting of massless scattering in radially symmetric shocks. Let
\be\label{B-defs}
\mathsf{A}_k=\frac{p_+}{\mu}\left(\frac{\sfv_k}{\sfv_k^2}-\frac{\sfv_{k-1}}{\sfv_{k-1}^2}\right):=\frac{p_+}{\mu}\,\mathsf{B}_{k}\,,
\ee
and define the effective potential
\be\label{eff-V-def-v-space}
    \mathbb{V}(\sft^2) := \mu^2\, \left|w(\mu^2 \sft^2)\right|^2 \;\qquad\text{obeying}\qquad \int\!{\hat\ud}^2\sft\, \mathbb{V}(\sft^2) = 1 \;.
\ee
In terms of these quantities, the radiation density is
\be\label{rad-dens-sym1}
s\,\frac{\ud^2\langle\!\langle\mathbb{P}_{1\to2}\rangle\!\rangle}{\ud x^{-}\,\ud s}=\frac{2\,\alpha}{\pi^2\,\lambda}\,(1-s)\,\Re\,\int\ud^2\sfv\sum_{n=0}^{\infty}\left[\prod_{j=1}^{n+2}\hat{\ud}^2\sft_j\,\mathbb{V}(\sft_j^2)\right]\mathsf{B}_1\cdot\mathsf{B}_{n+2}\left(\left[\prod_{k=1}^{n+1}\frac{1}{1-\im\,\theta_k}\right]-1\right)\,,
\ee
with $\theta_k$ given by \eqref{theta-k-radial}.

Changing integration variables from $\sfv$ to $\sfv_1$ and then re-labelling all indices, \eqref{rad-dens-sym1} becomes
\begin{multline}\label{rad-dens-sym2}
s\,\frac{\ud^2\langle\!\langle\mathbb{P}_{1\to2}\rangle\!\rangle}{\ud x^{-}\,\ud s}=\frac{2\,\alpha}{\pi^2\,\lambda}\,(1-s)\,\Re\,\int\ud^2\sfv\,\hat{\ud}^2\sft\,\mathbb{V}(\sft^2) \\
\times\,\mathsf{B}_0\cdot\,\sum_{n=0}^{\infty}\left[\prod_{j=1}^{n+1}\hat{\ud}^2\sft_j\,\mathbb{V}(\sft_j^2)\right]\mathsf{B}_{n+1}\left(\left[\prod_{k=0}^{n}\frac{1}{1-\im\,\theta_k}\right]-1\right)\;,
\end{multline}
in which
\be
    \mathsf{B}_0 := \frac{\sfv}{\sfv^2} - \frac{\sfv-\sft}{(\sfv-\sft)^2} \;.
\ee
By defining
\be\label{Baier-pattern-source}
    \sfJ_0(\sfv) := \int \hat{\ud}^2\sft\, \mathbb{V}(\sft^2)\,\mathsf{B}_0  = \frac{\sfv}{\sfv^2} \int \hat{\ud}^2\sft\,\Theta(\sft^2-\sfv^2)\,\mathbb{V}(\sft^2) \;,
\ee
(where the second equality follows from radial symmetry and the explicit form of $\mathsf{B}_0$) and
\be\label{radial-f-def}
\sff_{\kappa}(\sfv):=\sum_{n=0}^{\infty}\left[\prod_{j=1}^{n+1}\int\hat{\ud}^2\sft_j\,\mathbb{V}(\sft^2_j)\right]\mathsf{B}_{n+1}\,\prod_{k=0}^{n}\frac{1}{1-\im\,\theta_k}\,,
\ee
the radiation density can be written compactly as:
\be\label{Baier-pattern-spectrum}
s\frac{\ud^2 \langle\!\langle \mathbb{P}_{1\to2}\rangle\!\rangle}{\ud x^\LCm\,\ud s}
     = 
     \frac{2\alpha}{\pi\,\lambda}\,(1-s)\, \Re\!\int\!\ud^2\sfv\, \mathsf{J}_0(\sfv)\cdot \left(\sff_\kappa(\sfv)-\sff_0(\sfv)\right)\,.
\ee
The utility of writing the radiation density in this way is that the vector-valued function $\sff_{\kappa}(\sfv)$ can be seen, from its definition \eqref{radial-f-def}, to obey the integral equation
\be\label{Baier-pattern-recurrence}
    \left(1- \im\, \kappa\, \sfv^2\right) \sff_\kappa(\sfv) =
    -\sfJ_0(\sfv)
    + \int{\hat \ud}^2\sft\, \mathbb{V}(\sft^2)\, \sff_\kappa(\sfv+\sft) \;.
\ee
Thus, determining the spectrum boils down to solving this integral equation in the appropriate regime.

Consequently, it will be useful to remark on the behaviour of $\sff_{\kappa}(\sfv)$ in some universal limits. In particular, consider the large $|\sfv|$ behaviour of this function. In the $|\sfv|\to\infty$ limit, in \eqref{Baier-pattern-recurrence} one can approximate
\be\label{radsym-largev1}
\int{\hat \ud}^2\sft\, \mathbb{V}(\sft^2)\, \sff_\kappa(\sfv+\sft)\simeq \sff_{\kappa}(\sfv)\,\int{\hat \ud}^2\sft\, \mathbb{V}(\sft^2)=\sff_{\kappa}(\sfv)\,,
\ee
using the normalisation \eqref{eff-V-def-v-space}. Combined with \eqref{Baier-pattern-source}, this means that
\be\label{radsym-largev2}
\lim_{|\sfv|\to\infty}\sff_{\kappa}(\sfv)\simeq-\frac{\im\,\sfv}{\kappa\,\sfv^4}\int\hat{\ud}^2\sft\,\Theta(\sft^2-\sfv^2)\,\mathbb{V}(\sft^2)\,,
\ee
asymptotically. In other words, we see that $\sff_{\kappa}$ decays as $|\sfv|$ grows very large. This effectively provides a boundary condition for solving \eqref{Baier-pattern-recurrence}.

In addition, one can read off the solution for $\kappa=0$ by inspection. Indeed, note that
\be\label{radsymk=0}
\sff_{0}(\sfv)=-\frac{\sfv}{\sfv^2}\,,
\ee
obeys
\be\label{radsymk=0.1}
\begin{split}
-\mathsf{J}_0(\sfv)+\int\hat{\ud}^2\sft\,\mathbb{V}(\sft^2)\,\sff_0(\sfv+\sft) &=-\frac{\sfv}{\sfv^2}\int\hat{\ud}^2\sft\,\Theta(\sft^2-\sfv^2)\,\mathbb{V}(\sft^2)-\int\hat{\ud}^2\sft\,\mathbb{V}(\sft^2)\,\frac{\sfv+\sft}{(\sfv+\sft)^2} \\
&= -\frac{\sfv}{\sfv^2}\int\hat{\ud}^2\sft\,\Theta(\sft^2-\sfv^2)\,\mathbb{V}(\sft^2)-\frac{\sfv}{\sfv^2}\int\hat{\ud}^2\sft\,\Theta(\sfv^2-\sft^2)\,\mathbb{V}(\sft^2) \\
&=-\frac{\sfv}{\sfv^2}\int\hat{\ud}^2\sft\,\mathbb{V}(\sft^2)=-\frac{\sfv}{\sfv^2}=\sff_0(\sfv)\,,
\end{split}
\ee
as required. Note also that this particular solution is consistent with the large-$|\sfv|$ behaviour obtained for general $\kappa$ in \eqref{radsym-largev2}.

\medskip

Expressions (\ref{Baier-pattern-spectrum}) -- (\ref{Baier-pattern-recurrence}) are identical in form to those appearing in the literature for high-energy QED scattering through a medium modelled by screened Coulomb charges; our notation has been deliberately chosen to mimic the formulae appearing in~\cite{Baier:1996vi}. However, there are some key distinctions which are worth pointing out. Firstly, literature results make use of high-energy probe and soft-collinear photon approximations, whereas our results allow for more general kinematics; in terms of the remaining variables this means arbitrary $s\in(0,1)$ and arbitrary $p_+$, in particular allowing us to go beyond the soft and collinear limits. Secondly, the potential $\mathbb{V}$ underlying \eqref{Baier-pattern-spectrum} is \emph{all-orders} in the coupling between the probe and the background, whereas that in~\cite{Baier:1996vi} is quadratic in the coupling (i.e., it holds at leading perturbative order).

\subsection{Bethe-Heitler limit $\kappa\gg 1$}\label{BH-radial-section}
To analyze the large $\kappa$ limit, we will use a different method than~\cite{Baier:1996vi}, one which is more easily generalised to other cases. To simplify the integral equation \eqref{Baier-pattern-recurrence} in the $\kappa\gg1$ limit, change variables $\sfv \to \sfv/\sqrt{\kappa}$, and define $\sff$ by $\sff_\kappa(\sfv/\sqrt{\kappa}) := \sqrt{\kappa}\,\sff(\sfv)$. With this, all terms in \eqref{Baier-pattern-recurrence} pick up a factor of $\sqrt{\kappa}$, which can be divided out. The source term on the right hand side of \eqref{Baier-pattern-recurrence} then becomes
    \be
        -\mathsf{J}_0(\sfv)=-\frac{\sfv}{\sfv^2} \int\!\hat{\ud}^2\sft\,\Theta(\sft^2-\kappa^{-1}\sfv^2)\,\mathbb{V}(\sft^2) \underset{\kappa\gg1}{\simeq}
                -\frac{\sfv}{\sfv^2} \,,
    \ee
after using the normalisation condition \eqref{eff-V-def-v-space}. Meanwhile, the homogenous term on the right hand side of \eqref{Baier-pattern-recurrence} contains (an integral of) $\sff(\sfv+\sqrt{\kappa}\sft)$. From the asymptotics \eqref{radsym-largev2}, we know that $\sff$ vanishes for large argument, so as $\kappa\gg1$ this term can be neglected. One then obtains, in the large $\kappa$ regime, the approximate solution of the integral equation \eqref{Baier-pattern-recurrence} as
    \be\label{radsym-BH-f}
        \sff(\sfv)\underset{\kappa\gg1}{\simeq} - \frac{1}{1-\im\,\sfv^2}\,\frac{\sfv}{\sfv^2} \;.
    \ee
Naively, this makes it seem that the spectrum \eqref{Baier-pattern-spectrum} is now independent of $\kappa$, but care must be taken: the approximations used to obtain \eqref{radsym-BH-f} are clearly violated when the (rescaled) transverse variables $\sfv$ are such that $\sfv^2>\kappa$, in which case the spectral integral as given is divergent. However, for large $|\sfv|$, we have established in \eqref{radsym-largev2} that contributions from $\sff$ are negligible for any value of $\kappa$.

Consequently, the spectral integrals in this $\kappa\gg1$ approximation should only be performed up to $|\sfv|\sim\sqrt{\kappa}$. Implementing this in \eqref{Baier-pattern-spectrum} and using \eqref{radsymk=0}, one obtains the radiation density in the $\kappa\gg1$ regime as:
\be\label{radsym-BH1}
\begin{split}
\left.s\,\frac{\ud^2 \langle\!\langle \mathbb{P}_{1\to2}\rangle\!\rangle}{\ud x^\LCm\,\ud s}\right|^{\mathrm{BH}}_{\kappa\gg1}&=\frac{2\,\alpha}{\pi^2\,\lambda}\,(1-s)\int\ud^2\sfv\,
\frac{\sfv^2}{1+\sfv^4}\,\Theta(\kappa-\sfv^2) \\
& = \frac{\alpha}{\pi\,\lambda}\,(1-s)\,\log(1+\kappa^2) \\
& \simeq \frac{2\,\alpha}{\pi\,\lambda}\,(1-s)\,\log\kappa\,.
\end{split}
\ee
As expected, this produces the famous Bethe-Heitler result~\cite{Bethe:1934za} for the regime of independent radiation, which is independent of the details of the effective potential $\mathbb{V}(\sft^2)$ or shock profiles.

\subsection{LPM limit $\kappa\ll1$}

We now turn to the opposite regime, where $\kappa$ is very small. In this case, the analysis proceeds along the same lines as in~\cite{Baier:1996vi}, with the only differences being that our $\kappa$ allows for more general kinematics and the effective potential $\mathbb{V}$ \eqref{big-V-def} is all-orders, rather than linear. We will provide a brief review of the procedure here, but refer the reader to~\cite{Baier:1996vi} for futher details.

When $\kappa\ll1$, it is convenient to Fourier transform the integral equation \eqref{Baier-pattern-recurrence} from the transverse momentum variables $\sfv$ to transverse impact parameters $\sfb$; this results in a differential (rather than integral) equation:
\be\label{rad-LPM1}
    \left(1+ \im\, \kappa\, \nabla_\sfb^2\right) \sff_\kappa(\sfb) = -2\pi\im\, \frac{\sfb}{\sfb^2}\,(1-\tilde{\mathbb{V}}(\sfb^2)) + \tilde{\mathbb{V}}(\sfb^2)\,\sff_\kappa(\sfb) \;.
\ee
Here $\sff_{\kappa}(\sfb)$ is the Fourier transform of $\sff_{\kappa}(\sfv)$ and

\be\label{big-V-def}
    \tilde{\mathbb{V}}(\sfb^2):= \int{\ud}^2\sft\, \mathbb{V}(\sft^2)\, \e^{-\im\,\sfb\cdot\sft}  \;, 
\ee
is the effective potential in impact parameter space. Using the assumption of radial symmetry, this vectorial differential equation can be re-written as a scalar differential equation by setting
\be\label{rad-LPM2}
\sff_{\kappa}(\sfb)=\frac{\sfb}{\sfb^2}\left[h_{\kappa}(\sfb^2)-2\pi\im\right]\,.
\ee
Then \eqref{rad-LPM1} becomes
\be\label{rad-LPM3}
4\im\,\kappa\,h_{\kappa}''(\sfb^2)=\frac{(\tilde{\mathbb{V}}(\sfb^2)-1)}{\sfb^2}\,h_{\kappa}(\sfb^2)\,, \qquad h_{\kappa}''(\sfb^2)\equiv\frac{\ud^2 h_{\kappa}(\sfb^2)}{(\ud\sfb^2)^2}\,.
\ee
From this, we immediately deduce that $h_{0}(\sfb^2)=0$, while the boundary behaviour of $\sff_{\kappa}(\sfv)$ in momentum space implies the boundary conditions
\be\label{rad-LPM4}
h_{\kappa}(0)=2\pi\im\,, \qquad \lim_{\sfb^2\to\infty}h_{\kappa}(\sfb^2)=0\,,
\ee
in impact parameter space.

Armed with these facts, it is straightforward to re-write the radiation density in terms of an impact parameter integral: \be\label{rad-IP}
\begin{split}
s\,\frac{\ud^2 \langle\!\langle \mathbb{P}_{1\to2}\rangle\!\rangle}{\ud x^\LCm\,\ud s}& =\frac{2\,\alpha}{\pi\,\lambda}\,(1-s)\,\Re \int\frac{\ud r}{2\pi\,\im}\,\frac{(1-\tilde{\mathbb{V}}(r))}{r}\,h_{\kappa}(r) \\
&=\frac{4\,\alpha\,\kappa}{\pi^2\,\lambda}\,\Re\, h'_{\kappa}(0)\,.
\end{split}
\ee
At this point, this representation of the spectrum is valid for arbitrary $\kappa$, but requires one to solve for $h_{\kappa}(\sfb^2)$ explicitly, at least in the first formal neighbourhood of the origin. 

When $\kappa\ll1$, the differential equation \eqref{rad-LPM3} can be solved using a WKB approximation (with $\kappa$ playing the role of $\hbar^2$) to obtain~\cite{Baier:1996vi}:
\be\label{rad-LPM5}
h_{\kappa\ll1}(\sfb^2)=2\pi\im\left[\frac{-\tilde{\mathbb{V}}'(\sfb^2)\,\sfb^2}{1-\tilde{\mathbb{V}}(\sfb^2)}\right]^{\frac{1}{4}}\exp\!\left[-\sqrt{\frac{\im}{4\,\kappa}}\,\int_{0}^{\sfb^2}\ud r\,\sqrt{\frac{1-\tilde{\mathbb{V}}(r)}{r}}\right]\,.
\ee
Feeding this into \eqref{rad-IP}, one obtains
\be\label{rad-LPM-gen}
\left.s\,\frac{\ud^2 \langle\!\langle \mathbb{P}_{1\to2}\rangle\!\rangle}{\ud x^\LCm\,\ud s}\right|^{\mathrm{LPM}}_{\kappa\ll1}=\frac{2\,\alpha}{\pi\,\lambda}\,\sqrt{-2\,\kappa\,\tilde{\mathbb{V}}'(0)}\,,
\ee
which coincides with the functional form of Migdal's original result~\cite{Migdal:1956tc}. The power-law dependence on $\kappa$ should be compared to the logarithmic dependence exhibited in the BH regime (\ref{radsym-BH1}) and indicates a suppression of the radiation due to destructive interference in the LPM regime. 

\medskip

This result assumes, of course, that $\tilde{\mathbb{V}}'(0)$ is well-defined; it is easy to see that this is \emph{not} the case in the prototypical example of radially-symmetric shocks corresponding to the ultraboost of a Coulomb potential. Indeed, consider the shock profiles given by \eqref{Coulombshock}, $\Phi(|\sfX|)=(Q/2\pi)\log(\mu\,|\sfX|)$, where $\mu$ is an arbitrary mass scale. In this case~\cite{tHooft:1987vrq,Jackiw:1991ck},
\be\label{Coul-1}
W(\sfc)=\int\ud^2\sfX\,\exp\!\left[-\im\,\sfc\cdot\sfX+\frac{\im\,e\,Q}{2\,\pi}\,\log(\mu\,|\sfX|)\right]=\frac{\im\,Q}{\sfc^2}\left(\frac{4\,\mu^2}{\sfc^2}\right)^{\frac{\im\,eQ}{2\pi}}\frac{\Gamma\!\left(\frac{\im\,eQ}{4\pi}\right)}{\Gamma\!\left(-\frac{\im\,eQ}{4\pi}\right)}\,,
\ee
and while this clearly contains all-order information in $e$ and $Q$, its modulus-squared is equivalent to that obtained from scattering of a linear Coulomb potential
\be\label{Coul-2}
|W(\sfc)|^2=\frac{Q^2}{\sfc^4}\,.
\ee
In addition to normalisation by the infinite volume of the transverse plane as in \eqref{needed-norm}, one must also account for the IR divergence associated with integrating a Fourier-transformed Coulomb potential. The IR divergences can be regulated in the usual way with a mass scale $\mu$ (which can be freely chosen to be the same mass scale required to define the shock profile) as $\sfc^4\to(\sfc^2+\mu^2)^2$. The normalised transverse distribution is then fixed by requiring that \eqref{reg-dist2} is obeyed, leading to
\be\label{Coul-3}
|w(\sfc)|^{2}=\frac{1}{\pi}\,\frac{\mu^2}{(\sfc^2+\mu^2)^2}\,,
\ee
for the Coulombic shock profiles. Surprisingly, all dependence on the shock charge $Q$ is washed out by the normalisation condition\footnote{The same is true in the case of static charges, see~\cite{Baier:1996vi}.}, and all integrals weighted by this distribution behave as
\be\label{Coul-4}
\int\!{\hat\ud}^2\sfc\, |w(\sfc)|^2\, F(\sfc) = \frac{1}{\pi}  \int\!{\hat\ud}^2\sft\, \frac{1}{(\sft^2+1)^2}\, F(\mu \sft) \,,
\ee
for any test function $F$. Consequently, the effective potential in impact parameter space is given by
\be\label{Coul-5}
\tilde{\mathbb{V}}(\sfb^2)=\int\frac{{\ud}^2\sft}{\pi}\,\frac{\e^{-\im\,\sfb\cdot\sft}}{(\sft^2+1)^2}=|\sfb|\,K_{1}(\sfb)\,,
\ee
where $K_1$ is the modified Bessel function of the second kind.

This can now be fed into the differential equation \eqref{rad-LPM3} and evaluated in the small-$\sfb^2$ limit, since we are only interested in the solution near $\sfb^2=0$, which leads to the approximate differential equation~\cite{Chadan:1996rz}
\be\label{Coul-6}
h_{\kappa}''(\sfb^2)\simeq \frac{1}{16\im\,\kappa}\,\log(\sfb^{2})\,h_{\kappa}(\sfb^2)\,,
\ee
that can in turn be solved for small values of $\kappa$ using the boundary conditions \eqref{rad-LPM4}. The result is~\cite{Baier:1996vi}
\be\label{rad-LPM-Coul}
\left.s\,\frac{\ud^2 \langle\!\langle \mathbb{P}_{1\to2}\rangle\!\rangle}{\ud x^\LCm\,\ud s}\right|^{\mathrm{LPM,}\,\mathrm{Coulomb}}_{\kappa\ll1}=\frac{\alpha}{\pi\,\lambda}\,\sqrt{\kappa\,\log\!\left(\frac{1}{\kappa}\right)}\,.
\ee
With this explicit choice of potential, the suppression of the radiation due to destructive interference in the LPM regime relative to the BH regime \eqref{radsym-BH1} is clear.

\section{Impulsive plane waves}
\label{sec:IPW}

While the case of radially symmetric, sourced shocks led to the reproduction of well-known results for the BH and LPM regimes of the spectrum with some modest improvements in generality, the source-free case has received substantially less attention in the literature. In this case, we are interested in the doubly-averaged radiation density produced by a probe scattering of multiple \emph{impulsive plane waves} (IPWs), whose profiles take the form \eqref{IPW}. While this scenario no longer has the interpretation as a model for high-energy scattering through a medium (as was the case for sourced shocks), it \emph{is} a model for probe scattering through a train of ultrashort waves, of the sort that may be encountered in high-intensity laser-electron collisions.

With profile $e\,\Phi(\sfX)=\mathsf{a}\cdot\sfX$, it follows that the normalised transverse momentum distribution for the IPW is simply
\be
    |w(\sfc)|^2 = {\hat\delta}^2_\epsilon(\sfc-\mathsf{a}) \;,
\ee
where $\delta^2_{\epsilon}$ is any normalized regularization of the two-dimensional delta function with regularization parameter $\epsilon$ (e.g., $\delta^2_{\epsilon}(\sfX)=\frac{\e^{-\sfX^2/2\epsilon}}{2\pi\epsilon}$). The details of this regularization are completely irrelevant, since in the radiation density \eqref{rad-dens2} one can immediately take $\epsilon\to0$ and perform all ${\hat\ud}^2\sfc_j$ integrals against the resulting delta functions. Consequently, the doubly-averaged radiation density associated with scattering through a large sequence of IPWs is given by
\be\label{rad-densIPW}
s\,\frac{\ud^2 \langle\!\langle \mathbb{P}_{1\to2}\rangle\!\rangle}{\ud x^\LCm\,\ud s}=\frac{2\,\alpha}{\pi^2\,\lambda}\,(1-s)\,\Re\,\int\ud^2\sfu\,\sum_{n=0}^{\infty}\mathsf{C}_1\cdot\mathsf{C}_{n+2}\left(\left[\prod_{k=1}^{n+1}\frac{1}{1-\im\,\theta_k}\right]-1\right)\,,
\ee
where
\be\label{IPW-C}
\mathsf{C}_k:=\frac{\sfu+k\,\frac{\mathsf{a}}{p_+}}{\left(\sfu+k\,\frac{\mathsf{a}}{p_+}\right)^2+\frac{m^2}{p^2_+}}-\frac{\sfu+(k-1)\,\frac{\mathsf{a}}{p_+}}{\left(\sfu+(k-1)\,\frac{\mathsf{a}}{p_+}\right)^2+\frac{m^2}{p^2_+}}\,,
\ee
and the phases are now
\be\label{IPW-theta}
\theta_k=\frac{\lambda\,p_+}{2}\,\frac{s}{(1-s)}\left[\left(\sfu+k\,\frac{\mathsf{a}}{p_+}\right)^2+\frac{m^2}{p^2_+}\right]\,.
\ee
Further evaluation of the observable is both qualitatively and quantitatively different depending on whether the probe is taken to be massive or massless. We proceed to consider each case separately.

\subsection{Massive probes}

For massive probes, the spectrum naturally depends on \emph{two} dimensionless combinations of the photon momentum fraction $s$ and the other collision parameters: 
\be\label{gamma-kappa}
    \gamma = \frac{\lambda\, m^2}{2\,p_\LCp}\,\frac{s}{1-s} \;,
    \qquad
    \kappa = \frac{\lambda\, \mathsf{a}^2}{2\,p_\LCp}\,\frac{s}{1-s} \;,
\ee
related by $\gamma=m^2\,\kappa/\mathsf{a}^2$. Changing transverse momentum variables from $\sfu$ to
\be\label{IPW-vdef}
    \sfv:=\frac{p_+\,\sqrt{\gamma}}{m}\,\sfu \;, 
\ee
and setting
\be\label{IPW-hdef}
\sfh:=\frac{\sqrt{\gamma}}{m}\,\mathsf{a}=\sqrt{\kappa}\,\frac{\mathsf{a}}{|\mathsf{a}|}\,,
\ee
it follows that the accumulated phases \eqref{IPW-theta} can be rewritten compactly as
\be\label{IPW-theta2}
\theta_{k}=\gamma\left[1+\gamma^{-1}\,(\sfv+k\,\sfh)^2\right]\,.
\ee
The radiation density itself is then given by
\be\label{mIPW1}
s\,\frac{\ud^2 \langle\!\langle \mathbb{P}_{1\to2}\rangle\!\rangle}{\ud x^\LCm\,\ud s}=\frac{2\,\alpha}{\pi^2\,\lambda}\,(1-s)\,\Re\,\int\frac{\ud^2\sfv}{\gamma^2}\,\sum_{n=0}^{\infty}\mathsf{J}_1\cdot\mathsf{J}_{n+2}\left(\left[\prod_{k=1}^{n+1}\frac{1}{1-\im\,\theta_k}\right]-1\right)\,,
\ee
where the re-scaled quantities $\mathsf{J}_k$ are defined, for $k\geq 0$, by 
\be\label{mIPW-Jdef}
\mathsf{J}_{k}:=\frac{m\,\sqrt{\gamma}}{p_+}\,\mathsf{C}_k=\frac{\sfv+k\,\sfh}{1+\gamma^{-1}(\sfv+k\,\sfh)^2}-\frac{\sfv+(k-1)\,\sfh}{1+\gamma^{-1}(\sfv+(k-1)\,\sfh)^2}\,.
\ee
As in the sourced case, it is now possible to further manipulate \eqref{mIPW1} to cast it in a form controlled by a master function solving an appropriate equation. Indeed, after changing integration variables from $\sfv$ to $\sfv+\sfh$ and then re-labelling dummy indices in the sums and products, \eqref{mIPW1} becomes
\be\label{mIPW2}
s\,\frac{\ud^2 \langle\!\langle \mathbb{P}_{1\to2}\rangle\!\rangle}{\ud x^\LCm\,\ud s}=\frac{2\,\alpha}{\pi^2\,\lambda}\,(1-s)\,\Re\,\int\frac{\ud^2\sfv}{\gamma^2}\,\mathsf{J}_0\cdot\left(\sff(\sfv)-\sff_0(\sfv)\right)\,,
\ee
with the functions
\be\label{IPW-f}
\sff(\sfv):=
\sum_{n=0}^{\infty}\mathsf{J}_{n+1}\,\left[\prod_{k=0}^{n}
\frac{1}{1-\im\,\theta_k}\right] \;, \qquad
\sff_0(\sfv):= - \frac{\sfv}{1+\gamma^{-1}\sfv^2} \;.
\ee
From its definition, it can easily be established that $\sff(\sfv)$ obeys:
\be\label{keep-massive-recurrence}
\begin{split}
    \left(1-\im\,\gamma-\im\,\sfv^2\right)\sff(\sfv) & = \frac{\sfv+\sfh}{1+\gamma^{-1}\,(\sfv+\sfh)^2}-\frac{\sfv}{1+\gamma^{-1}\sfv^2} + \sff(\sfv+\sfh) \\
    &=\mathsf{J}_1(\sfv)+\sff(\sfv+\sfh)\,.
\end{split}
\ee
Thus, we see that in the case of multiple IPWs, the integral equation \eqref{Baier-pattern-recurrence} becomes an algebraic recurrence relation. Using this recurrence relation, the expression for the radiation density can be further simplified to 
\be\label{to-do-reduced}
s\, \frac{\ud^2 \langle\!\langle \mathbb{P}_{1\to2}\rangle\!\rangle}{\ud s\,\ud x^\LCm}
    = \frac{2\,\alpha}{\pi^2\,\lambda}\,
    (1-s)\,\Re\,\int \ud^2\sfv\, \frac{\im}{\gamma}\, \sfv \cdot \sff(\sfv) \;,
\ee
and the entire problem is boiled down to solving for $\sff(\sfv)$.

Observe that (both here and in the literature), it is $\sqrt{\kappa}$ which appears in this recurrence relation as the appropriate expansion parameter -- recall that the transverse vector $\sfh\sim\sqrt{\kappa}$. This means that, crucially, the Bethe-Heitler and LPM limits are controlled by large and small $\kappa$, respectively: while the parameter $\gamma$ may also grow or shrink in these regimes, it is assumed that $\kappa$ is always the largest or smallest parameter. In other words, for the BH limit of $\kappa\gg1$, we additionally assume that $\kappa\gg\gamma$, or equivalently that $\mathsf{a}^2/m^2\gg1$. Similarly, for the LPM limit of $\kappa\ll1$, we assume that $\kappa\ll\gamma$ or $\mathsf{a}^2/m^2\ll1$. These are the strong and weak field regimes respectively~\cite{Ilderton:2019vot}, see also the discussion below in Section~\ref{subsec:comparison}.

Furthermore, the simplification of an algebraic recurrence relation over an integral equation means that \eqref{keep-massive-recurrence} can be solved \emph{exactly}, with the boundary condition that $\sff(\sfv)\to0$ as $|\sfv|\to\infty$ asymptotically. However, the resulting analytic expression is so lengthy that it is unwieldy for practical calculation of the spectrum -- see Appendix~\ref{app:exact}, where we give some details and also present the exact solution for the case of massless probes, which is more compact.


Without explicitly solving the recurrence relation, it is still possible to extract the asymptotic behaviour of $\sff(\sfv)$ (as we did in the case of radially symmetric shocks, see~\eqref{radsym-largev1}--\eqref{radsym-largev2}). As $|\sfv|\to\infty$, it follows that $\sff(\sfv+\sfh)\simeq\sff(\sfv)$, and the source terms corresponding to $\mathsf{J}_{1}$ in \eqref{keep-massive-recurrence} can be expanded to leading order in $|\sfv|^{-1}$. From this, one identifies
\be\label{f-asymptotics}
    \lim_{|\sfv|\to\infty}\sff(\sfv) \simeq \frac{\im\,\gamma}{\sfv^4}\left(\sfh -\frac{2\,\sfh\cdot\sfv}{\sfv^2}\,\sfv\right) \;,
\ee
demonstrating that $|\sff(\sfv)|$ falls rapidly as $1/\sfv^4$ when $|\sfv|\to\infty$.

\subsubsection{Bethe-Heitler limit $\kappa\gg 1$}
The large $\kappa$ regime is, in the parametrization of \eqref{IPW-hdef}, equivalent to determining $\sff(\sfv)$ when the recursion relation is evaluated at large $|\sfh|$. Now, the asymptotics \eqref{f-asymptotics} ensure that $\sff$ vanishes for large arguments, so when $\kappa\gg1$ it follows that one can approximate $\sff(\sfv+\sfh)\simeq0$, and only the $\sfh$-independent term in $\mathsf{J}_1$ contributes in the recurrence relation \eqref{keep-massive-recurrence}. From this we read off the approximate solution
\be\label{large-kappa-approx}
    \sff(\sfv) \underset{\kappa\gg1}{\simeq} \sff_{\text{\tiny BH}}(\sfv) := \frac{-\sfv}{(1-\im\,\gamma-\im\,\sfv^2)(1+\gamma^{-1}\,\sfv^2)} \;.
\ee
In the emission spectrum \eqref{to-do-reduced} one integrates over all $\sfv$, and clearly the approximation \eqref{large-kappa-approx} breaks down when $|\sfv|>|\sfh|=\sqrt{\kappa}$. However, invoking the general asymptotic behaviour \eqref{f-asymptotics}, it folows that the large $|\sfv|$ contribution to the integrand in \eqref{to-do-reduced} goes like $\sfv\cdot\sfh/\sfv^4$ and thus vanishes when integrated over angles. 

The radiation density is then approximated by
\be\begin{split}\label{spectrum-at-large-kappa}
\left.s\, \frac{\ud^2 \langle\!\langle \mathbb{P}_{1\to2}\rangle\!\rangle}{\ud s\,\ud x^\LCm}\right|^{\mathrm{BH}}_{\kappa\gg1}
    &= \frac{2\,\alpha}{\pi^2\,\lambda}\,
    (1-s)\,\Re\int\!\ud^2\sfv\, \Theta(\kappa-\sfv^2)\,\frac{\im}{\gamma}\, \sfv \cdot \sff_{\text{\tiny BH}}(\sfv) \\
    &\simeq  \frac{2\alpha}{\pi\lambda}
    (1-s) \log \kappa \;.
\end{split}
\ee
We obtain the \emph{same} $\log\kappa$ Bethe-Heitler scaling of the radiation density as in the sourced case \eqref{radsym-BH1}; this is consistent with the universality of the regime of independent radiation from each shock. We will comment further on the physical interpretation of the large (and small) $\kappa$ regimes below.

\subsubsection{LPM limit $\kappa\ll 1$}
%
For the opposite limit of small $\kappa$, it is convenient to introduce a vector-valued function $\sfg(\sfv)$ which is the (appropriately scaled) deviation of $\sff(\sfv)$ from the free solution:
\be\label{IPW-gdef}
\sfg(\sfv):=\sff(\sfv)+\frac{\sfv}{1+\gamma^{-1}\,\sfv^2}\,.
\ee
Written in terms of this $\sfg$, the recurrence relation simplifies to
\be\label{keep-massive-recurrence-g}
    \left(1-\im\,\gamma-\im\,\sfv^2\right)\sfg(\sfv) +\im\, \gamma\,\sfv  =\sfg(\sfv+\sfh) \;.
\ee
Note that because $\sff$ and $\sfg$ differ by a real function, either can be used to compute the emission spectrum \eqref{to-do-reduced}.

Now, since small $\kappa$ corresponds to small $\sfh$ we can expand \eqref{keep-massive-recurrence-g} to linear order in $\sfh$, which converts the algebraic recurrence relation into the differential equation
\be\label{g-DE}
    \sfh\cdot\nabla\sfg(\sfv) \underset{\kappa\ll1}{\simeq} \im\,\gamma\, \sfv - \im\,(\gamma+\sfv^2)\,\sfg(\sfv) \;.
\ee
At first sight, it might appear that by converting an algebraic relation into a differential equation, we are losing the advantage of considering source-free rather than sourced solutions. However, this is a linear, first-order equation which is far easier to solve than the original recurrence relation.

Without loss of generality, we can choose coordinates on the transverse plane such that $\sfh\equiv \sqrt{\kappa}(1,0)$. The solution  to \eqref{g-DE} which vanishes asymptotically is then
\be
    \sfg(v_1,v_2) =-\frac{\im\,\gamma}{\sqrt{\kappa}}\,
    \e^{-\im\, P(v_1,v_2)}\,\int_{v_1}^\infty\!\ud u\, \begin{pmatrix}u\\v_2\end{pmatrix} \e^{\im\, P(u,v_2)}\,,  
\ee
where
\be
    P(v_1,v_2) = \frac{1}{\sqrt{\kappa}} \left[ \left(\gamma+v_2^2\right) v_1 + \frac{v_1^3}{3}\right] \,.
\ee
The spectrum is then controlled by the integral:
\begin{multline}\label{g-int1}
\Re\,\int\ud^2\sfv\,\frac{\im}{\gamma}\,\sfv\cdot\sfg(\sfv) \\
=\frac{1}{\sqrt{\kappa}}\int\ud^{2}\sfv\int_{v_1}^{\infty}\ud u\left(v_1\,u+v_2^2\right)\left[\cos P(v_1,v_2)\,\cos P(u,v_2)+\sin P(v_1,v_2)\,\sin P(u,v_2)\right]\,.
\end{multline}
Observe that, making the changes of variables $u\to-u$ and $v_1\to-v_1$ in these integrals, we may write (\ref{g-int1}) as
\begin{multline}\label{g-int2}
(\ref{g-int1}) 
=\frac{1}{\sqrt{\kappa}}
\int\ud^{2}\sfv
\int^{v_1}_{-\infty}\ud u\left(v_1\,u+v_2^2\right)\left[\cos P(v_1,v_2)\,\cos P(u,v_2)\right. \\
\left.+\sin P(v_1,v_2)\,\sin P(u,v_2)\right]\,.
\end{multline}
It follows that the integral over $u$ in (\ref{g-int1}) can be extended to the entire real line, which allows us to drop several terms which integrate to zero (e.g., odd functions integrated over the real line), and we obtain
\begin{multline}\label{g-int3}
\Re\,\int\ud^2\sfv\,\frac{\im}{\gamma}\,\sfv\cdot\sfg(\sfv) \\
=\frac{1}{2\sqrt{\kappa}}\int\ud^{2}\sfv\int_{-\infty}^{\infty}\!\ud u\left[v_2^2\,\cos P(v_1,v_2)\,\cos P(u,v_2)+v_1\,u\,\sin P(v_1,v_2)\,\sin P(u,v_2)\right]\,,
\end{multline}
In this form, the $v_1$ and $u$ integrals have factorised and can be performed independently, with each returning Airy functions or their derivatives:
\be\label{prod-of-airy}
   \Re
    \int\!\ud^2\sfv\, \frac{\im}{\gamma}\,\sfv\cdot\sfg(\sfv) =\frac{(2\pi)^2}{2\,\sqrt{\kappa}}\int\!\ud v_2\left[\kappa^{1/3}\, v_2^2\, \mathrm{Ai}^2\!\left(\frac{\gamma+v_2^2}{\kappa^{1/3}}\right) + \kappa^{2/3}\,\mathrm{Ai}^{\prime\, 2}\!\left(\frac{\gamma+v_2^2}{\kappa^{1/3}}\right)\right]\,.
\ee
To evaluate the remaining integral over $v_2$, observe that $\gamma+v_2^2$ is bounded from below (by $\gamma$), so that for sufficiently small $\kappa$, as appropriate to the LPM regime, the Airy functions appearing in \eqref{prod-of-airy} can be replaced by their asymptotic expansions~\cite[\href{https://https://dlmf.nist.gov/9.7}{(9.7)}]{NIST:DLMF}
\be
    \mathrm{Ai}(x) \sim \frac{\e^{-\tfrac23 x^{3/2}}}{2 \sqrt{\pi}\, x^{1/4}} \;,
    \qquad
    \mathrm{Ai}'(x) \sim -\frac{\e^{-\tfrac23 x^{3/2}} x^{1/4}}{2 \sqrt{\pi}} \;,
\ee
resulting in
\be\begin{split}
    \Re\!
    \int\!\ud^2\sfv\, \frac{i}{\gamma}\sfv.\sfg(\sfv) &\simeq \frac{\pi}{2} \int\!\ud v_2\, \frac{(\gamma +2v_2^2) }{ \sqrt{\gamma +v_2^2}}\,\, \exp\bigg[-\frac{4}{3\sqrt{\kappa}} (\gamma +v_2^2)^{3/2}\bigg] \;.
\end{split}
\ee
The integral may now be performed using Laplace's method, expanding around the minimum of the exponent at $v_2=0$.  Finally, we arrive at
\be\label{final-massive}
\left.s\, \frac{\ud^2 \langle\!\langle \mathbb{P}_{1\to2}\rangle\!\rangle}{\ud s\,\ud x^\LCm}\right|^{\mathrm{LPM}}_{\kappa\ll1}
    = \frac{\alpha\,(1-s)}{\lambda\,\sqrt{2\pi}}
    \, (\gamma\kappa)^{1/4}\, 
    \exp\!\left[-\frac{4}{3}\frac{\gamma ^{3/2}}{\kappa^{1/2}}\right] \;,
\ee
for the LPM regime of the radiation density. 

In comparison with the logarithmic behaviour at large $\kappa$, there is very strong exponential suppression of the spectrum, with a non-perturbative dependence on $\kappa$. This suppression is also substantially stronger than for the `standard' LPM effect with radially symmetric sources \eqref{rad-LPM-gen}, \eqref{rad-LPM-Coul}. Observe that the form of the result is somewhat tidier when considering the differential probability, rather than the radiation density:
\be\label{final-massive-rewritten}
\frac{\ud^2 \langle\!\langle \mathbb{P}_{1\to2}\rangle\!\rangle}{\ud s\,\ud x^\LCm}
    = \frac{\alpha}{2\,\sqrt{2\pi}}\,
    \frac{m^2}{p_\LCp}
    \,
    \frac{\kappa^{1/4}}{\gamma^{3/4}}\, 
    \exp\!\left[-\frac{4}{3}\frac{\gamma ^{3/2}}{\kappa^{1/2}}\right]\,.
\ee
As expected, the differential probability is infrared divergent, but the exponential suppression and non-perturbative dependence on $\kappa$ remains.

\subsection{Massless probes}

The case of a massless probe particle is distinguished first of all by the fact that only a single dimensionless parameter is in play: rather than the $\gamma$ and $\kappa$ of \eqref{gamma-kappa}, we only have $\kappa$. Setting $m=0$ in the general expression \eqref{rad-densIPW}, one can change transverse variables from $\sfu$ to
\be\label{ml-v}
\sfv:=\sfu\,\sqrt{\frac{\lambda\,s\,p_+}{2\,(1-s)}}\,,
\ee
under which the radiation density becomes
\be\label{0IPW1}
s\,\frac{\ud^2 \langle\!\langle \mathbb{P}_{1\to2}\rangle\!\rangle}{\ud x^\LCm\,\ud s}=\frac{2\,\alpha}{\pi^2\,\lambda}\,(1-s)\,\Re\,\int\ud^2\sfv\left(\frac{\sfv}{\sfv^2}-\frac{\sfv-\sfh}{(\sfv-\sfh)^2}\right)\cdot\sff(\sfv)\,,
\ee
where $\sfh:=\sqrt{\kappa}\mathsf{a}/|\mathsf{a}|$ and now
\be\label{0f-def}
\sff(\sfv):=\sum_{k=0}^\infty\left(\frac{\sfv+(k+1)\,\sfh}{(\sfv+(k+1)\,\sfh)^2}-\frac{\sfv+k\,\sfh}{(\sfv+k\,\sfh)^2}\right)\prod_{j=0}^{k}\left[1-\im\,(\sfv+j\,\sfh)^2\right]^{-1}\,.
\ee
Once again, the function $\sff$ obeys a recurrence relation
\be\label{0f-recur}
\left(1-\im\,\sfv^2\right)\sff(\sfv) = \frac{\sfv+\sfh}{(\sfv+\sfh)^2} - \frac{\sfv}{\sfv^2} + \sff(\sfv+\sfh)\,,
\ee
which is simplified by working with $\sfg=\sff+\sfv/\sfv^2$:
\be\label{massless-g-recurrence}
    \left(1-\im\,\sfv^2\right)\sfg(\sfv) + \im\, \sfv = \sfg(\sfv+\sfh) \;.
\ee
This allows the spectrum to be written as
\be\label{massless-spectrum-simplified}
s\, \frac{\ud^2 \langle\!\langle \mathbb{P}_{1\to2}\rangle\!\rangle}{\ud s\,\ud x^\LCm}
    = \frac{2\,\alpha}{\pi^2\,\lambda}\,
    (1-s)\,\Re\,\int\ud^2\sfv\, \im\, \sfv \cdot \sfg(\sfv) \;,
\ee
entirely in terms of $\sfg(\sfv)$. These expressions are similar to those of the massive case, although simpler: the only parameter appearing in the recurrence relation is $\kappa$, and there are no additional factors in the spectrum \eqref{massless-spectrum-simplified}, as opposed to \eqref{to-do-reduced}.

\medskip

We can now proceed to analyse the radiation density in the regimes of large and small $\kappa$, following similar steps to the massive case. For $\kappa\gg1$, this follows exactly the same lines as before, leading to an identical result \eqref{spectrum-at-large-kappa} for the spectrum. This is consistent, as the earlier result was actually independent of the probe mass, and again reinforces the universality of the Bethe-Heitler regime of independent radiation.

Turning to the $\kappa\ll1$ limit, we follow the same approach as before, expanding \eqref{massless-g-recurrence} to obtain a first-order differential equation for $\sfg(\sfv)$. The solution is similar -- but not identical -- to that found in the massive case: 
\be
    \sfg(v_1,v_2) =\frac{-\im}{\sqrt{\kappa}}\,
    \e^{-\im\, P(v_1,v_2)}\int_{v_1}^\infty\!\ud u\, \begin{pmatrix}u\\v_2\end{pmatrix} \e^{\im\, P(u,v_2)}   \;, \quad 
    P(v_1,v_2) = \frac{1}{\sqrt{\kappa}} \Big(\frac13 v_1^3+ v_2^2 v_1\Big) \;.
\ee
Inserting this into \eqref{massless-spectrum-simplified}, one quickly arrives at the equivalent of \eqref{prod-of-airy}:
\be
\begin{split}\label{almost-there}
\Re
    \int\!\ud^2\sfv\, \im\,\sfv\cdot\sfg(\sfv)
 = \frac{(2\pi)^2}{\sqrt{\kappa}}\int\!\ud v_2\left[ \kappa^{1/3}\, v_2^2\, \mathrm{Ai}^2\!\left(\frac{v_2^2}{\kappa^{1/3}}\right) + \kappa^{2/3}\,\mathrm{Ai}^{\prime\, 2}\!\left(\frac{v_2^2}{\kappa^{1/3}}\right)\right] \,.
\end{split}
\ee
It is here that the difference between the massless and massive cases becomes clear: it is no longer possible to use the asymptotic expansion of the Airy functions, as there is nothing remaining in their arguments to set a scale. 

Indeed, observe that changing variables $v_2 = \kappa^{1/6}\,x$ removes $\kappa$ from the integral and establishes that the radiation density must scale like $\kappa^{1/3}$ overall. It is also possible to compute the Airy integrals \eqref{almost-there} to obtain the constant of proportionality exactly. Using~\cite[\href{https://https://dlmf.nist.gov/9.11}{(9.11)}]{NIST:DLMF},
\be
    \int\limits_{-\infty}^\infty\!\ud x \left[ x^2\, \textrm{Ai}^2(x) + \textrm{Ai}^{\prime\, 2}(x)\right]
    =
    \frac{\left(\frac{2}{3}\right)^{\tfrac13}}{\Gamma \left(\frac{1}{3}\right)} \;,
\ee
we arrive at the final result
\be
\begin{split}\label{final-massless}
\left.s\, \frac{\ud^2 \langle\!\langle \mathbb{P}_{1\to 2}\rangle\!\rangle}{\ud s\,\ud x^\LCm}\right|^{\mathrm{LPM}}_{\kappa\ll1,\,m=0}
&=      \frac{4\,\alpha}{\lambda} \left(\frac{2}{3}\right)^{1/3}\,(1-s)\,  \frac{\kappa^{1/3}}{\Gamma\!\left(\frac{1}{3}\right)}\,.
\end{split}
\ee
As for the case of massive probes, we find a suppression of the spectrum at small $\kappa$, but this time with power-law, rather than exponential, dependence. The difference in functional form of the suppression is due to the mass which, through $\gamma$, introduces a scale and provides a lower bound on the interference phases $\theta_k$ entering the action. This emphasises the differences between massive and massless theories. It is nevertheless interesting to observe that, despite the differences in their asymptotic forms, the limits $\kappa\to 0$ and $m\to 0$ do strictly commute, with both yielding zero for the radiation density.

\subsection{Comparison across regimes and field models}\label{subsec:comparison}
We end this section by setting our results for multiple scattering on IPWs into the wider context of particle scattering strong-field QED. In the study of intense laser-matter interactions, it is typical to describe the laser fields as plane waves. Within this model there is almost arbitrary freedom in choosing a (temporal) pulse shape and duration for the electromagnetic fields. The two `extreme' cases which allow for the most analytic control are the impulsive plane wave (IPW) limit considered above, and the opposite limit of a `constant crossed field' (CCF), with potential 
\be\label{ccf-potential}
    A_\mu(x) = - n_\mu\, x^\LCperp E_\LCperp \;,
\ee
in which $E_\LCperp$ is a constant (transverse) vector -- the only nonzero field strength component is $F_{\LCm\LCperp} = E_\LCperp$, constant.  The CCF is often adopted as a first approximation of the strong-field limit of other electromagnetic fields, in particular those of an ultra-intense laser\footnote{More accurately, the CCF is extrapolated to a `locally constant field approximation' which can handle more accurate descriptions of focussed laser field geometry. The model has several known shortcomings, see~\cite{Harvey:2014qla,DiPiazza:2017raw,Ilderton:2018nws,Fedotov:2022ely}, one of which we will revisit below.}. A CCF can be thought of as the zero-frequency limit of a general plane wave, and is homogeneous in space and time. This should be contrasted against the IPW, which is completely localised in lightfront time and thus contains all frequencies in its Fourier transform. Our goal in this subsection is to explore the similarities and differences between scattering on multiple IPWs and on a CCF.

There are many different physical regimes of light-matter interaction. A useful parameter with which to characterise characterising these is 
    \be
        \xi = \frac{eE}{m\,\omega} \;,
    \ee
in which $E$ and $\omega$ typify, respectively, the electric field strength and frequency of the wave.
This `dimensionless intensity parameter' $\xi$ quantifies the strength of the background, and is the effective coupling constant to matter. In the strong-field regime $\xi>1$ the interactions between probes and waves  must be treated non-perturbatively (e.g., through the background field methods we have employed here). The energy of interaction, which has the same null geometry as for the IPW, is parameterised by $\eta=\omega n\cdot p/m^2 = \omega p_\LCp/m^2$, essentially the centre of mass energy. Combining these intensity and energy parameters gives
    \be\label{chi-def}
    \chi := \xi\, \eta 
    \frac{eE\, p_\LCp}{m^3} \;,
\ee
which quantifies strong-field quantum effects~\cite{Fedotov:2022ely}. (We consider the massless case separately, below.)

Of particular interest are the (differences between) the low and high $\chi$ regimes~\cite{Fedotov:2016afw}, where one may find non-perturbative dependence on $\chi$ or strong field phenomena including pair production and vacuum polarisation effects -- for reviews, see~\cite{Gonoskov:2021hwf,Fedotov:2022ely}.
Furthermore, $\chi$ can clearly be made large by going to high field strengths, high energy, or both -- which is chosen leads to different behaviours in physical processes~\cite{Podszus:2018hnz,Ilderton:2019kqp}. We will therefore ask how the small and large $\chi$ limits translate into the language of the LPM effect (following which we will easily be able to compare with literature results for scattering on a CCF).  

For an impulsive plane wave, the electric field strength is not defined, but the \emph{work done} by the field on a charged particle is. For a particle of charge $e$, this work is $|\mathsf{a}|$, or $N|\mathsf{a}|$ for $N$ impulsive waves~\cite{Ilderton:2019vot}. Averaging provides a total length scale of $N\lambda$, hence we can define an effective (constant) electric field
\be\label{E-eff-def}
    eE_{\text{eff}} := \frac{N|\mathsf{a}|}{N\lambda} = \frac{|\mathsf{a}|}{\lambda} \;.
\ee
From this we obtain, using the definition \eqref{chi-def}, an effective $\chi=|\mathsf{a}|\, p_\LCp/(\lambda m^3)$.
It is immediately clear, comparing with the parameter definitions \eqref{gamma-kappa}, that the limits of large or small $\chi$ project onto the limits of large or small $\kappa$, respectively, since
\be
    \kappa = \frac{\lambda\, \mathsf{a}^2}{2\,p_\LCp}\,\frac{s}{1-s}
    =
    \chi^2\left(\frac{\lambda\, m^2}{p_\LCp}\right)^3\,\frac{s}{1-s}\,.
\ee
Hence, we may simply take our previous results for the BH and LPM limits, \eqref{spectrum-at-large-kappa} and \eqref{final-massive} respectively, and rewrite them in terms of $\chi$ to obtain:
\be\label{summary-results-in-chi-IPW}
   s\, \frac{\ud^2 \langle\!\langle \mathbb{P}_{1\to2}\rangle\!\rangle}{\ud s\,\ud x^\LCm}
=
\begin{cases}
\displaystyle\frac{4\alpha}{\pi\lambda}\,
    (1-s) \log \chi \;, & \chi \to \infty \;, \\[20pt]
\displaystyle      
{\frac12}
{\frac{\alpha\, m^2}{\sqrt{\pi}\,p_\LCp}\,\sqrt{s(1-s)}\,\sqrt{\chi}\,\,
\exp\!\bigg[-\frac{2}{3\chi}\frac{s}{1-s}\bigg]} \;, & \chi \to 0 \;.
    \end{cases}
\ee
We see that the LPM suppression of radiation emerges in the small $\chi$ limit, and exhibits a non-perturbative dependence on $\chi$.  This is not unprecedented -- for instance, similar exponents appear in the probability of pair production at low $\chi$~\cite{Ritus:1985vta,King:2024ffy}.
The BH limit, on the other hand, is the large $\chi$ limit. The logarithmic behaviour seen here is consistent with other results in impulsive fields~\cite{Ilderton:2019vot}.

With this reparameterisation of our results in hand, we are ready to compare with literature expressions for the radiation density of a scalar particle scattering on a CCF with potential (\ref{ccf-potential}). 

\medskip

\paragraph{The massive case:} Particle scattering on a CCF is analysed by constructing the appropriate external wavefunctions, inserting into (\ref{three-point-amp-start}) to compute the emission amplitude, and mod-squaring, all as we did for the IPW. In this case, though, there no averaging is required. The calculation is well-studied in the literature (cf.,~\cite{Ritus:1985vta,Fedotov:2022ely}), so we proceed directly to the final expression for the radiation density. This depends non-trivially only on $\chi$ (not $\xi$ or $\eta$) as there is no frequency in a CCF. The radiation density may be written, using the same notation as above~\cite{Ritus:1985vta}, as
\be\label{ccf-integral}
    s\,\frac{\ud^2 {\mathbb P}_{\mathrm{CCF}}}{\ud s\, \ud x^\LCm} = -\frac{2\alpha}{\pi}\,\frac{m^2}{p_\LCp}\,s \int\limits_0^\infty\!\ud\tau\, \tau\, \sin\left(z\tau+\frac{\tau^3}{3}\right)\bigg[\frac{1}{3z+\tau^2}-\frac{1}{z}\bigg] \;,
\ee
in which
\be
 z := \bigg(\frac{1}{\chi}\frac{s}{1-s}\bigg)^{2/3} \;.
\ee
It is interesting to note that the integration variable $\tau$ is itself a kind of interference phase: it arises at the level of the probability as the difference in lightfront times of the interaction vertex in the amplitude and in it conjugate. The integral may be performed in terms of Airy functions, but we will not need the exact expression, as we focus only on the two limits of large and small $\chi$. 

The large $\chi$ limit is easily taken by expanding \eqref{ccf-integral} in small $z$ and performing the resulting $z$-independent integral. For the small $\chi$ limit, one expands the factor in square brackets in large $z$; the $\tau$-integral then gives an Airy function, and this can be approximated by its standard asymptotic limit. We obtain
\be\label{summary-results-in-chi-CCF}
    s\,\frac{\ud^2 {\mathbb P}_{\mathrm{CCF}}}{\ud s\, \ud x^\LCm} = 
    \begin{cases}
    \displaystyle
      3^{\tfrac16}\,\Gamma\!\left(\tfrac{2}{3}\right)
      \,
      \frac{\alpha\, m^2}{\pi\,  p_\LCp} \big(s (1 -s)^2\big)^{1/3}
      \,
      \chi^{2/3} \;, & \chi \to \infty \;,\\[20pt]
\displaystyle
{\frac{2}{3}}\,
{\frac{\alpha\,  m^2 }{\sqrt{\pi}\, p_\LCp} 
\sqrt{s(1-s)}\,\sqrt{\chi}}\,\,
\exp\!\bigg[-\frac{2}{3\chi}\frac{s}{1-s}\bigg] \;, & \chi \to 0 \;.
    \end{cases}
\ee
There are notable similarities and differences compared to \eqref{summary-results-in-chi-IPW}. Perhaps most striking is that, at small $\chi$, we see the same non-perturbative dependence; indeed, the \emph{only} difference between the $\chi\to0$ radiation densities for a CFF and multiple IPWs is an overall numerical prefactor.
Otherwise the small $\chi$ limits of (\ref{summary-results-in-chi-IPW}) and (\ref{summary-results-in-chi-CCF}) are, as functions of $s$ and $\chi$, completely equivalent. Qualitatively, one might say that a very large stack of $N$ closely packed, ultrashort but otherwise identical waves can look like a `block' of homogeneous field.

For large $\chi$, on the other hand, the radiation density is very different in the two field configurations. There is a power law scaling in the CCF, in stark contrast to the logarithmic scaling in IPWs (and in the static case typically studied in the literature) -- this is physically reasonable since, in the large $\chi$ limit, a sequence of well-separated waves does \emph{not} look like a block of homogeneous field.

We can also comment briefly on the $s\to 0$ behaviour of our results. Since $s=\ell_\LCp/p_\LCp$, this limit corresponds to either the frequency of the outgoing photon vanishing, or to the photon being emitted in the same null direction $n_\mu$ in which the wave propagates. It is well-known that the CCF model of a laser fails to correctly capture the small-$s$ behaviour found otherwise universally in QED: the emission probability should go like $\ud\mathbb{P}/\ud s \sim 1/s$, but instead goes like $\sim s^{-2/3}$~\cite{Harvey:2014qla,DiPiazza:2017raw,Ilderton:2018nws}. This is simply because the genuinely constant field does not admit scattering boundary conditions, and hence the IR physics is different -- as soon as a finite pulse duration is introduced, the expected soft behaviour returns~\cite{Dinu:2012tj}. 
    
The similarity of the IPW and CCF results in the small $\chi$ limit then suggests that `edge effects' have been lost from the IPW results at some stage, and indeed it is easy to track down where. Observe that the terms dropped between \eqref{incoh-sum} and \eqref{largeN} when taking the large $N$ limit include those which carry the usual soft divergence. Retaining `finite size' effects in our calculations is therefore a potential avenue for future study, see also Sec.~\ref{sec:concs} below.

\medskip

\paragraph{The massless case:} Finally, we compare the radiation density in the CCF and IPW models for  \emph{massless} probes. In this case there is no $\chi$, since there is no mass, but $\kappa$ is still proportional to the effective field strength (\ref{E-eff-def}) squared. The strong and weak field limits can then still be read off directly from the BH and LPM limits.
The former exhibits the universal logarithmic behaviour, while the latter is given by \eqref{final-massless}; in terms of \eqref{E-eff-def} these become
\be\label{massless-summary}
   s\, \frac{\ud^2 \langle\!\langle \mathbb{P}_{1\to2}\rangle\!\rangle}{\ud s\,\ud x^\LCm}\bigg|_{m=0}
=
\begin{cases}
\displaystyle       \frac{4\alpha}{\pi\lambda}\,
    (1-s)\, \log \frac{eE_{\text{eff}}}{p_\LCp^2}  & \kappa \to \infty \;, \\[20pt]
\displaystyle      
3^{\tfrac16}\,\Gamma\!\left(\tfrac{2}{3}\right)
      \,
      {\frac{2\alpha\, p_\LCp}{\pi}} \big(s (1 -s)^2\big)^{1/3}\,\,
      {\left(\frac{eE_{\text{eff}}}{p_\LCp^2}\right)^{2/3}} & \kappa \to 0 \;,
    \end{cases}
\ee
keeping only the field-strength dependence under the log in the BH limit.

In the case of scattering on a CCF, the massless probe limit is obtained by taking $m\to 0$ under the integral in \eqref{ccf-integral}. The integral may then be computed without further approximation and one obtains, for any field strength,
\be\label{massless-ccf-final}
    s\,\frac{\ud^2 {\mathbb P}_{\mathrm{CCF}}}{\ud s\, \ud x^\LCm}\bigg|_{m=0}
    =
    3^{\tfrac16}\,\Gamma\!\left(\tfrac{2}{3}\right)
      \,
      {\frac{\alpha\, p_\LCp}{\pi}} \big(s (1 -s)^2\big)^{1/3}\,\,
      {\left(\frac{eE}{p_\LCp^2}\right)^{2/3}} \;.
\ee
As for massive probes, we see very different behaviours in the BH limit of scattering on a CCF compared to scattering on multiple shocks \eqref{massless-summary}, but agreement in the LPM limit\footnote{The LPM limits for massive and massless probes, (\ref{summary-results-in-chi-CCF}) and (\ref{massless-ccf-final}), are in fact the same -- the mass drops out of the former. Thus the massless limit seems to project onto the large $\chi$ limit. This may have some practical use when discussing the `Ritus-Narozhny conjecture', see~\cite{Fedotov:2016afw}, which we will consider elsewhere.}, up to numerical factors.

\section{Conclusions}
\label{sec:concs}
In this paper, we investigated the radiation produced by scattering a (scalar) electron off multiple null shockwaves, corresponding to ultraboosted charge distributions and impulsive (ultrashort) electromagnetic waves.
We began by computing the amplitudes for elastic scattering and for photon emission on the multi-shock background. These amplitudes for sourced shocks are always weighted averages over the corresponding amplitudes in source-free impulsive plane waves, generalising an observation made previously in the case of scattering on a single shock~\cite{Adamo:2021jxz,Ilderton:2024ufp}.
Taking the modulus-squared of the photon emission amplitude and averaging over the shock locations, we computed the photon emission specrum and radiation density. We then analysed this in two classes of background field (i.e., for two classes of shock profile in the transverse plane) and in two physical regimes.

Our first example was the class of shock profiles with radial symmetry in the transverse plane -- this includes the ultraboosted Coulomb field of a point charge.
We investigated the Bethe-Heitler regime of radiation, characterised by $\kappa\gg 0$ (with $\kappa$ proportional to the shock separation and average momentum transfer), in which contributions to the emission spectrum from each shock are independent (essentially contributing \emph{incoherently}), and recovered the expected logarithmic dependence on $\kappa$. Turning to the small $\kappa$ regime, we observed that the spectrum is suppressed due to destructive interference between emissions from different shocks, recovering the famous Landau-Pomeranchuk-Migdal (LPM) effect. Functionally, our results match exactly with the standard LPM spectrum in this case (cf.,~\cite{Baier:1996vi}), but now with general scattering kinematics and arbitrary boosted charge sources thanks to our strong-field QED setup.

In our second example we considered scattering on multiple impulsive plane waves -- source free solutions of Maxwell's equations of ultrashort duration. The resulting radiation spectrum again exhibited logarithmic behaviour in the Bethe-Heitler regime. Indeed, this is universal, the result being independent of the shock profile and probe particle mass. In the LPM regime, the radiation spectrum again showed suppression, relative to the BH regime, but the functional form of this suppression was qualitatively different for massive and massless probes. In the massless case we saw power-law suppression, while for massive particles the suppression was exponential, and in fact \emph{non-perturbative} in the controlling parameter $\kappa$. 

In general, we were able to write the radiation density as an integral, over transverse momentum transfer, of a vector-function $\sff_\kappa$ defined via an integro-differential equation \eqref{Baier-pattern-recurrence}. Solving this equation was the central (technical) obstacle in constructing closed-form results for the radiation density. Literature calculations for radial shocks rely, in the LPM regime, on converting this equation to a second-order differential equation in impact-parameter space and using WKB methods. In the case of scattering on multiple impulsive waves, though, the defining integral relation becomes an (algebraic) recurrence relation, and our method of solution was therefore different. For the LPM lmit we remained in momentum space (as we did for the BH limit of both the wave and radial shock backgrounds), and showed how to turn the recurrence relation into a \emph{first}-order differential equation, whose solution leads to an expression for the radiation density in terms of integrals of Airy functions. We were then able to evaluate these integrals using known asymptotics of the Airy function, leading to non-perturbative suppression of radiation.

\medskip

There are various extensions of our approach which provide opportunities for future work. The approximations we have made ultimately treat the sequence of (a large number of) shocks as an infinite medium. A natural extension is then to reintroduce finite size effects into the calculations, allowing us to model e.g.~scattering on thin targets~\cite{Blankenbecler:1995dy,Zakharov:1996cm,Baier:1997wa}. This would be especially interesting given the connections we observed between our results and scattering on a constant crossed field (CCF). This model of an intense laser field also ignores edge effects (as do the `locally constant' simulation models developed from it), resulting in infra-red behaviour which is at odds with the expected universal soft behaviour of QED, among other issues~\cite{Dinu:2012tj,Harvey:2014qla,DiPiazza:2017raw,Ilderton:2018nws,Fedotov:2022ely,Montefiori:2026bit}.

A major advantage of formulating the multiple scattering problem in terms of null shocks is the ability to generalise to other theories. Such shockwaves are easily extended to solutions of non-abelian gauge theory and gravity~\cite{Bonnor:1969mfs,Aichelburg:1970dh,Trautman:1980bj}, and their null nature ensures that they are also solutions of any effective field theory (EFT) with Yang-Mills or general relativity as its low-energy limit~\cite{Amati:1988sa,Amati:1988tn,Horowitz:1989bv}. In the context of Yang-Mills theory, non-abelian shockwaves are the bedrock of the colour-glass condensate, an EFT describing the high-energy, high-density regime of quantum chromodynamics (QCD)~\cite{Balitsky:1995ub,Iancu:2003xm,Gelis:2010nm}, while shockwaves are ubiquitous tools for studying eikonal resummation, causality and universality in quantum gravity~\cite{tHooft:1987vrq,Amati:1988tn,Kabat:1992tb,Camanho:2014apa,Kologlu:2019bco,DiVecchia:2023frv}.

Of course, the study of gluonic radiation from multiple scatterings and the LPM effect in QCD are very well-studied~(c.f., \cite{Baier:2000mf}). However, just as we were able to uncover some more general properties of these phenomena in QED with the exact background field theory treatment based on shockwaves, we expect that something similar should be possible in the non-abelian setting. In QCD, the major complication relative to QED will be the fact that emitted gluons are not free, but the Feynman rules for gluons in a shockwave background have been known for some time~\cite{McLerran:1994vd,Ayala:1995kg,Gyulassy:1997vt,Balitsky:1998ya}, so this is not a major technical obstacle. While the direct computation of gravitational radiation from multi-shock scattering will obviously be even more complicated, there is the possibility that this could be inferred from the QCD result using double copy. Indeed, double copy structures in the Regge limit of gravity go back to Lipatov~\cite{Lipatov:1982vv} and have been the subject of much recent study~\cite{Raj:2023iqn,Raj:2023irr,Raj:2024xsi,Rothstein:2024nlq,Raj:2025hse,Stasto:2026xqw}. A gravitational avatar of the LPM effect was recently observed in~\cite{Blackstad:2026eoa}; there, the effect seems to be associated with non-null corrections to the shockwave approximation, so it would be very interesting to see what -- if any -- portion of the effect is captured by our multi-shock setup.

\acknowledgments

It is a pleasure to thank Arseny Mironov and Raju Venugopalan for many interesting conversations and comments on the draft. The authors are supported by a Royal Society University Research Fellowship (TA), a Royal Society Summer Studentship (HK), the Simons Collaboration on Celestial Holography CH-00001550-11 (TA), the ERC Consolidator/UKRI Frontier grant TwistorQFT EP/Z000157/1 (TA), the STFC consolidated grant ST/X000494/1 (TA \& AI), and the Leverhulme grant RPG-2025-301 (AI). The Center of Gravity is a Center of Excellence funded by the Danish National Research Foundation under grant no.~DNRF184.

\appendix

\section{Exact solution of the recurrence relation}\label{app:exact}

The recurrence relation \eqref{keep-massive-recurrence} for the function governing the radiation density from IPWs can be solved exactly. Substituting the relation into itself, iterating, and imposing that $\sff(\sfv)\to 0$ as $|\sfv|\to \infty$, one finds
\be\label{f-exact-1}
    \sff(\sfv) = \sum_{k=0}^\infty \frac{\sfJ_1(\sfv + k \sfh)}{\prod_{j=0}^k d(\sfv + j\sfh)} \;,
\ee
where $\sfJ_1(\sfv)$ is the source term on the RHS of (\ref{keep-massive-recurrence}), and
\be
    d(\sfv) = 1-\im\,\gamma-\im\,\sfv^2 \;.
\ee
We now go to a frame where $\sfh\equiv \sqrt{\kappa}\,(1,0)$ and factorise $d(\sfv+j\sfh)$ as
    \be\label{zpm-def}
        d(\sfv+j\sfh) = -\im\,\kappa\, (j+z_\LCp)(j+z_\LCm) \quad\text{where}\quad
        z_\LCpm = \frac{v_1 \pm \im \sqrt{\im + v_2^2 + \gamma}}{\sqrt{\kappa}} \;.
    \ee
The denominator in \eqref{f-exact-1} can then be written in terms of the Gamma function as
\be\label{f-exact-2}
        \prod_{j=0}^k d(\sfv + j\sfh) = (-\im\,\kappa)^{k+1} \, \frac{\Gamma(z_\LCp+k+1)}{\Gamma(z_\LCp)}
        \frac{\Gamma(z_\LCm+k+1)}{\Gamma(z_\LCm)} \;,
\ee
and hence
\be\label{f-exact-3}
    \sff(\sfv) = \Gamma(z_\LCp)\Gamma(z_\LCm)\sum_{k=0}^\infty  \frac{\sfJ_1(\sfv + k \sfh)}{\Gamma(z_\LCp+k+1)\Gamma(z_\LCm+k+1)} \bigg(\frac{\im}{\kappa}\bigg)^{k+1}\;.
\ee
The remaining sum in \ref{f-exact-3} can be performed in terms of hypergeometric functions. However, the resulting expression is extremely lengthy and difficult to process, so we do not report it here. 

Instead we can illustrate the form of the full result by writing down the simpler solution of \eqref{massless-g-recurrence} for $\sfg(\sfv)$ in the case of a massless probe. In this context, we have
\be\label{massless-g-recurrence-reprise}
    \left(1-\im\,\sfv^2\right)\sfg(\sfv) + \im\, \sfv = \sfg(\sfv+\sfh) \implies \sfg(\sfv) = \sum\limits_{k=0}^\infty \frac{-\im\,(\sfv+k\,\sfh)}{\prod_{j=0}^{k} \big[1-\im\,(\sfv+j \sfh)^2\big]}\;.
\ee
The product in the denominator returns the same result as in \eqref{f-exact-2} but with $\gamma=0$ in the definitions of $z_\LCpm$, see \eqref{zpm-def}. The remaining sum can then be performed to yield
\be\label{g-massless-sol}
\begin{split}
\sfg(\sfv)
&=
- \im\, \sfv\, \frac{_1F_2\!\left(1;1+z_\LCm,1+z_\LCp;\tfrac{\im}{\kappa}\right)}{(1-\im\,\sfv^2)}
-\im\,\sfh\, \frac{\, _1F_2\!\left(2;2+z_\LCm,2+z_\LCp;\tfrac{\im}{\kappa}\right)}{(1-\im\,\sfv^2) (1-\im\,(\sfv+\sfh)^2)} \;,
\end{split}
\ee
in which $_1F_2$ is the generalised hypergeometric function~\cite[\href{https://https://dlmf.nist.gov/16}{(16)}]{NIST:DLMF}. One can see already in this expression the non-perturbative dependence of the result on $\kappa$, both explicitly and through $z_\LCpm$. In the massive case, the general solution is a far lengthier sum over hypergeometric functions, but with a similar non-perturbative dependence on $\kappa$, as well as various dependences on $z_\LCpm$ and~$\gamma$. 

\bibliographystyle{JHEP}

\begin{thebibliography}{10}

\bibitem{Ter-Mikaelian:1972}
M.~L. Ter-Mikaelian, \emph{{High Energy Electromagnetic Processes in Condensed
  Media}}.
\newblock John Wiley and Sons, 1972.

\bibitem{Akhiezer:1996pa}
A.~I. Akhiezer and N.~F. Shulga, \emph{{High-energy electrodynamics in
  matter}}.
\newblock Gordon and Breach, 1996.

\bibitem{Bethe:1934za}
H.~Bethe and W.~Heitler, \emph{{On the Stopping of fast particles and on the
  creation of positive electrons}},
  \href{http://dx.doi.org/10.1098/rspa.1934.0140}{\emph{Proc. Roy. Soc. Lond.
  A} {\bfseries 146} (1934) 83--112}.

\bibitem{Landau:1953um}
L.~D. Landau and I.~Pomeranchuk, \emph{{Limits of applicability of the theory
  of bremsstrahlung electrons and pair production at high-energies}},
  {\emph{Dokl. Akad. Nauk Ser. Fiz.} {\bfseries 92} (1953) 535--536}.

\bibitem{Landau:1953gr}
L.~D. Landau and I.~Pomeranchuk, \emph{{Electron cascade process at very
  high-energies}}, {\emph{Dokl. Akad. Nauk Ser. Fiz.} {\bfseries 92} (1953)
  735--738}.

\bibitem{Feinberg:1956}
E.~L. Feinberg and I.~Pomeranchuk, \emph{{High energy inelastic diffraction
  phenomena}}, \href{http://dx.doi.org/10.1007/BF02746068}{\emph{Nuovo Cim.}
  {\bfseries 3} (1956) 652–--671}.

\bibitem{Migdal:1956tc}
A.~B. Migdal, \emph{{Bremsstrahlung and Pair Production at High Energies in
  Condensed Media}},
  \href{http://dx.doi.org/10.1103/PhysRev.103.1811}{\emph{Phys. Rev.}
  {\bfseries 103} (1956) 1811--1820}.

\bibitem{Anthony:1995fs}
P.~L. Anthony et~al., \emph{{An Accurate measurement of the
  Landau-Pomeranchuk-Migdal effect}},
  \href{http://dx.doi.org/10.1103/PhysRevLett.75.1949}{\emph{Phys. Rev. Lett.}
  {\bfseries 75} (1995) 1949--1952}.

\bibitem{SLAC-E-146:1997hnd}
{\scshape SLAC-E-146} collaboration, P.~L. Anthony et~al.,
  \emph{{Bremsstrahlung suppression due to the LPM and dielectric effects in a
  variety of materials}},
  \href{http://dx.doi.org/10.1103/PhysRevD.56.1373}{\emph{Phys. Rev. D}
  {\bfseries 56} (1997) 1373--1390},
  [\href{https://arxiv.org/abs/hep-ex/9703016}{{\ttfamily hep-ex/9703016}}].

\bibitem{Stanev:1982au}
T.~Stanev, C.~Vankov, R.~E. Streitmatter, R.~W. Ellsworth and T.~Bowen,
  \emph{{Development of ultrahigh-energy electromagnetic cascades in water and
  lead including the Landau-Pomeranchuk-Migdal effect}},
  \href{http://dx.doi.org/10.1103/PhysRevD.25.1291}{\emph{Phys. Rev. D}
  {\bfseries 25} (1982) 1291--1304}.

\bibitem{Zas:1991jv}
E.~Zas, F.~Halzen and T.~Stanev, \emph{{Electromagnetic pulses from high-energy
  showers: Implications for neutrino detection}},
  \href{http://dx.doi.org/10.1103/PhysRevD.45.362}{\emph{Phys. Rev. D}
  {\bfseries 45} (1992) 362--376}.

\bibitem{Blumer:2009jrd}
J.~Blumer, R.~Engel and J.~R. Horandel, \emph{{Cosmic Rays from the Knee to the
  Highest Energies}},
  \href{http://dx.doi.org/10.1016/j.ppnp.2009.05.002}{\emph{Prog. Part. Nucl.
  Phys.} {\bfseries 63} (2009) 293--338},
  [\href{https://arxiv.org/abs/0904.0725}{{\ttfamily 0904.0725}}].

\bibitem{Gerhardt:2010bj}
L.~Gerhardt and S.~R. Klein, \emph{{Electron and Photon Interactions in the
  Regime of Strong LPM Suppression}},
  \href{http://dx.doi.org/10.1103/PhysRevD.82.074017}{\emph{Phys. Rev. D}
  {\bfseries 82} (2010) 074017},
  [\href{https://arxiv.org/abs/1007.0039}{{\ttfamily 1007.0039}}].

\bibitem{Anchordoqui:2018qom}
L.~A. Anchordoqui, \emph{{Ultra-High-Energy Cosmic Rays}},
  \href{http://dx.doi.org/10.1016/j.physrep.2019.01.002}{\emph{Phys. Rept.}
  {\bfseries 801} (2019) 1--93},
  [\href{https://arxiv.org/abs/1807.09645}{{\ttfamily 1807.09645}}].

\bibitem{Gyulassy:1993hr}
M.~Gyulassy and X.-n. Wang, \emph{{Multiple collisions and induced gluon
  Bremsstrahlung in QCD}},
  \href{http://dx.doi.org/10.1016/0550-3213(94)90079-5}{\emph{Nucl. Phys. B}
  {\bfseries 420} (1994) 583--614},
  [\href{https://arxiv.org/abs/nucl-th/9306003}{{\ttfamily nucl-th/9306003}}].

\bibitem{Wang:1994fx}
X.-N. Wang, M.~Gyulassy and M.~Plumer, \emph{{The LPM effect in QCD and
  radiative energy loss in a quark gluon plasma}},
  \href{http://dx.doi.org/10.1103/PhysRevD.51.3436}{\emph{Phys. Rev. D}
  {\bfseries 51} (1995) 3436--3446},
  [\href{https://arxiv.org/abs/hep-ph/9408344}{{\ttfamily hep-ph/9408344}}].

\bibitem{Perl:1994bg}
M.~L. Perl, \emph{{Notes on the Landau, Pomeranchuk, Migdal effect: Experiment
  and theory}},  in \emph{{8th Les Rencontres de Physique de la Vallee d'Aoste:
  Results and Perspectives in Particle Physics}}, pp.~463--477, 5, 1994.

\bibitem{Levin:1995mx}
E.~Levin, \emph{{Landau-Pomeranchuk-Migdal effect for nuclear matter in QCD}},
  \href{http://dx.doi.org/10.1016/0370-2693(96)00526-6}{\emph{Phys. Lett. B}
  {\bfseries 380} (1996) 399--408},
  [\href{https://arxiv.org/abs/hep-ph/9508414}{{\ttfamily hep-ph/9508414}}].

\bibitem{Baier:1996vi}
R.~Baier, Y.~L. Dokshitzer, A.~H. Mueller, S.~Peigne and D.~Schiff, \emph{{The
  Landau-Pomeranchuk-Migdal effect in QED}},
  \href{http://dx.doi.org/10.1016/0550-3213(96)00426-9}{\emph{Nucl. Phys. B}
  {\bfseries 478} (1996) 577--597},
  [\href{https://arxiv.org/abs/hep-ph/9604327}{{\ttfamily hep-ph/9604327}}].

\bibitem{Zakharov:1996fv}
B.~G. Zakharov, \emph{{Fully quantum treatment of the Landau-Pomeranchuk-Migdal
  effect in QED and QCD}}, \href{http://dx.doi.org/10.1134/1.567126}{\emph{JETP
  Lett.} {\bfseries 63} (1996) 952--957},
  [\href{https://arxiv.org/abs/hep-ph/9607440}{{\ttfamily hep-ph/9607440}}].

\bibitem{Zakharov:1998sv}
B.~G. Zakharov, \emph{{Light cone path integral approach to the
  Landau-Pomeranchuk-Migdal effect}}, {\emph{Phys. Atom. Nucl.} {\bfseries 61}
  (1998) 838--854}, [\href{https://arxiv.org/abs/hep-ph/9807540}{{\ttfamily
  hep-ph/9807540}}].

\bibitem{Baier:1998kq}
R.~Baier, Y.~L. Dokshitzer, A.~H. Mueller and D.~Schiff, \emph{{Medium induced
  radiative energy loss: Equivalence between the BDMPS and Zakharov
  formalisms}},
  \href{http://dx.doi.org/10.1016/S0550-3213(98)00546-X}{\emph{Nucl. Phys. B}
  {\bfseries 531} (1998) 403--425},
  [\href{https://arxiv.org/abs/hep-ph/9804212}{{\ttfamily hep-ph/9804212}}].

\bibitem{Klein:1998du}
S.~Klein, \emph{{Suppression of Bremsstrahlung and pair production due to
  environmental factors}},
  \href{http://dx.doi.org/10.1103/RevModPhys.71.1501}{\emph{Rev. Mod. Phys.}
  {\bfseries 71} (1999) 1501--1538},
  [\href{https://arxiv.org/abs/hep-ph/9802442}{{\ttfamily hep-ph/9802442}}].

\bibitem{Baier:2000mf}
R.~Baier, D.~Schiff and B.~G. Zakharov, \emph{{Energy loss in perturbative
  QCD}}, \href{http://dx.doi.org/10.1146/annurev.nucl.50.1.37}{\emph{Ann. Rev.
  Nucl. Part. Sci.} {\bfseries 50} (2000) 37--69},
  [\href{https://arxiv.org/abs/hep-ph/0002198}{{\ttfamily hep-ph/0002198}}].

\bibitem{Furry:1951bef}
W.~H. Furry, \emph{{On Bound States and Scattering in Positron Theory}},
  \href{http://dx.doi.org/10.1103/PhysRev.81.115}{\emph{Phys. Rev.} {\bfseries
  81} (1951) 115}.

\bibitem{Gonoskov:2021hwf}
A.~Gonoskov, T.~G. Blackburn, M.~Marklund and S.~S. Bulanov, \emph{{Charged
  particle motion and radiation in strong electromagnetic fields}},
  \href{http://dx.doi.org/10.1103/RevModPhys.94.045001}{\emph{Rev. Mod. Phys.}
  {\bfseries 94} (2022) 045001},
  [\href{https://arxiv.org/abs/2107.02161}{{\ttfamily 2107.02161}}].

\bibitem{Fedotov:2022ely}
A.~Fedotov, A.~Ilderton, F.~Karbstein, B.~King, D.~Seipt, H.~Taya et~al.,
  \emph{{Advances in QED with intense background fields}},
  \href{http://dx.doi.org/10.1016/j.physrep.2023.01.003}{\emph{Phys. Rept.}
  {\bfseries 1010} (2023) 1--138},
  [\href{https://arxiv.org/abs/2203.00019}{{\ttfamily 2203.00019}}].

\bibitem{Galitsky:1964}
V.~M. Galitsky and I.~I. Gurevich, \emph{{Coherence effects in
  ultra-relativistic electron bremsstrahlung.}},
  \href{http://dx.doi.org/10.1007/BF02733969}{\emph{Nuovo Cim.} {\bfseries 32}
  (1964) 396--407}.

\bibitem{Arnold:2025dqj}
P.~Arnold, J.~Bautista, O.~Elgedawy and S.~Iqbal, \emph{{Revisiting extremely
  high energy QED bremsstrahlung in matter: large modifications to the LPM
  effect}}, \href{http://dx.doi.org/10.1007/JHEP03(2026)015}{\emph{JHEP}
  {\bfseries 03} (2026) 015},
  [\href{https://arxiv.org/abs/2508.21120}{{\ttfamily 2508.21120}}].

\bibitem{Bonnor:1969rb}
W.~B. Bonnor, \emph{{Solutions of Maxwell's equations for charge moving with
  the speed of light}}, \href{http://dx.doi.org/10.1007/BF00670703}{\emph{Int.
  J. Theor. Phys.} {\bfseries 2} (1969) 373--379}.

\bibitem{Aichelburg:1970dh}
P.~C. Aichelburg and R.~U. Sexl, \emph{{On the Gravitational field of a
  massless particle}}, \href{http://dx.doi.org/10.1007/BF00758149}{\emph{Gen.
  Rel. Grav.} {\bfseries 2} (1971) 303--312}.

\bibitem{Robinson:1984}
I.~Robinson and K.~R\'{o}zga, \emph{{Lightlike contractions on Minkowski
  space-time}}, \href{http://dx.doi.org/10.1063/1.526189}{\emph{J. Math. Phys.}
  {\bfseries 25} (1984) 499--505}.

\bibitem{Dray:1984ha}
T.~Dray and G.~'t~Hooft, \emph{{The Gravitational Shock Wave of a Massless
  Particle}}, \href{http://dx.doi.org/10.1016/0550-3213(85)90525-5}{\emph{Nucl.
  Phys. B} {\bfseries 253} (1985) 173--188}.

\bibitem{Ferrari:1988cc}
V.~Ferrari, P.~Pendenza and G.~Veneziano, \emph{{Beamlike Gravitational Waves
  and Their Geodesics}}, \href{http://dx.doi.org/10.1007/BF00758938}{\emph{Gen.
  Rel. Grav.} {\bfseries 20} (1988) 1185--1191}.

\bibitem{Jackiw:1991ck}
R.~Jackiw, D.~N. Kabat and M.~Ortiz, \emph{{Electromagnetic fields of a
  massless particle and the eikonal}},
  \href{http://dx.doi.org/10.1016/0370-2693(92)90971-6}{\emph{Phys. Lett. B}
  {\bfseries 277} (1992) 148--152},
  [\href{https://arxiv.org/abs/hep-th/9112020}{{\ttfamily hep-th/9112020}}].

\bibitem{Penrose:1972xrn}
R.~Penrose, \emph{{The geometry of impulsive gravitational waves}},
  pp.~101--115.
\newblock Clarendon Press, 1972.

\bibitem{Penrose:1965rx}
R.~Penrose, \emph{{A Remarkable property of plane waves in general
  relativity}}, \href{http://dx.doi.org/10.1103/RevModPhys.37.215}{\emph{Rev.
  Mod. Phys.} {\bfseries 37} (1965) 215--220}.

\bibitem{Lodone:2009qe}
P.~Lodone and V.~S. Rychkov, \emph{{Radiation Problem in Transplanckian
  Scattering}},
  \href{http://dx.doi.org/10.1088/1126-6708/2009/12/036}{\emph{JHEP} {\bfseries
  12} (2009) 036}, [\href{https://arxiv.org/abs/0909.3519}{{\ttfamily
  0909.3519}}].

\bibitem{Adamo:2021jxz}
T.~Adamo, A.~Ilderton and A.~J. MacLeod, \emph{{Particle-beam scattering from
  strong-field QED}},
  \href{http://dx.doi.org/10.1103/PhysRevD.104.116013}{\emph{Phys. Rev. D}
  {\bfseries 104} (2021) 116013},
  [\href{https://arxiv.org/abs/2110.02567}{{\ttfamily 2110.02567}}].

\bibitem{Fedotov:2013uja}
A.~M. Fedotov and A.~A. Mironov, \emph{{Pair creation by collision of an
  intense laser pulse with a high-frequency photon beam}},
  \href{http://dx.doi.org/10.1103/PhysRevA.88.062110}{\emph{Phys. Rev. A}
  {\bfseries 88} (2013) 062110},
  [\href{https://arxiv.org/abs/1310.7258}{{\ttfamily 1310.7258}}].

\bibitem{Ilderton:2019vot}
A.~Ilderton, \emph{{Exact results for scattering on ultrashort plane wave
  backgrounds}},
  \href{http://dx.doi.org/10.1103/PhysRevD.100.125018}{\emph{Phys. Rev. D}
  {\bfseries 100} (2019) 125018},
  [\href{https://arxiv.org/abs/1909.02484}{{\ttfamily 1909.02484}}].

\bibitem{Weinberg:1965nx}
S.~Weinberg, \emph{{Infrared photons and gravitons}},
  \href{http://dx.doi.org/10.1103/PhysRev.140.B516}{\emph{Phys. Rev.}
  {\bfseries 140} (1965) B516--B524}.

\bibitem{Taub:1948zza}
A.~H. Taub, \emph{{Orbits of Charged Particles in Constant Fields}},
  \href{http://dx.doi.org/10.1103/PhysRev.73.786}{\emph{Phys. Rev.} {\bfseries
  73} (1948) 786--798}.

\bibitem{Sengupta:1949}
N.~D. Sengupta, \emph{{On the scattering of electromagnetic waves by free
  electron-1: classical theory}}, {\emph{Bulletin of the Calcutta Mathematical
  Society} {\bfseries 41} (1949) 187}.

\bibitem{Adamo:2017nia}
T.~Adamo, E.~Casali, L.~Mason and S.~Nekovar, \emph{{Scattering on plane waves
  and the double copy}},
  \href{http://dx.doi.org/10.1088/1361-6382/aa9961}{\emph{Class. Quant. Grav.}
  {\bfseries 35} (2018) 015004},
  [\href{https://arxiv.org/abs/1706.08925}{{\ttfamily 1706.08925}}].

\bibitem{Adamo:2022rmp}
T.~Adamo, A.~Cristofoli and A.~Ilderton, \emph{{Classical physics from
  amplitudes on curved backgrounds}},
  \href{http://dx.doi.org/10.1007/JHEP08(2022)281}{\emph{JHEP} {\bfseries 08}
  (2022) 281}, [\href{https://arxiv.org/abs/2203.13785}{{\ttfamily
  2203.13785}}].

\bibitem{Copinger:2024twl}
P.~Copinger, J.~P. Edwards, A.~Ilderton and K.~Rajeev, \emph{{All-multiplicity
  amplitudes in impulsive PP-waves from the worldline formalism}},
  \href{http://dx.doi.org/10.1007/JHEP09(2024)148}{\emph{JHEP} {\bfseries 09}
  (2024) 148}, [\href{https://arxiv.org/abs/2405.07385}{{\ttfamily
  2405.07385}}].

\bibitem{Ilderton:2024ufp}
A.~Ilderton and H.~Kingham, \emph{{Vacuum polarization effects in impulsive
  fields}}, \href{http://dx.doi.org/10.1103/PhysRevD.110.105022}{\emph{Phys.
  Rev. D} {\bfseries 110} (2024) 105022},
  [\href{https://arxiv.org/abs/2409.01372}{{\ttfamily 2409.01372}}].

\bibitem{Peskin:1995ev}
M.~E. Peskin and D.~V. Schroeder, \emph{{An Introduction to quantum field
  theory}}.
\newblock Addison-Wesley, Reading, USA, 1995,
  \href{http://dx.doi.org/10.1201/9780429503559}{10.1201/9780429503559}.

\bibitem{Berestetskii:1982qgu}
V.~B. Berestetskii, E.~M. Lifshitz and L.~P. Pitaevskii, \emph{{Quantum
  Electrodynamics}}, vol.~4 of \emph{Course of Theoretical Physics}.
\newblock Pergamon Press, Oxford, 1982.

\bibitem{Brodsky:1997de}
S.~J. Brodsky, H.-C. Pauli and S.~S. Pinsky, \emph{{Quantum chromodynamics and
  other field theories on the light cone}},
  \href{http://dx.doi.org/10.1016/S0370-1573(97)00089-6}{\emph{Phys. Rept.}
  {\bfseries 301} (1998) 299--486},
  [\href{https://arxiv.org/abs/hep-ph/9705477}{{\ttfamily hep-ph/9705477}}].

\bibitem{Heinzl:2000ht}
T.~Heinzl, \emph{{Light cone quantization: Foundations and applications}},
  \href{http://dx.doi.org/10.1007/3-540-45114-5_2}{\emph{Lect. Notes Phys.}
  {\bfseries 572} (2001) 55--142},
  [\href{https://arxiv.org/abs/hep-th/0008096}{{\ttfamily hep-th/0008096}}].

\bibitem{Hebenstreit:2009km}
F.~Hebenstreit, R.~Alkofer, G.~V. Dunne and H.~Gies, \emph{{Momentum signatures
  for Schwinger pair production in short laser pulses with sub-cycle
  structure}},
  \href{http://dx.doi.org/10.1103/PhysRevLett.102.150404}{\emph{Phys. Rev.
  Lett.} {\bfseries 102} (2009) 150404},
  [\href{https://arxiv.org/abs/0901.2631}{{\ttfamily 0901.2631}}].

\bibitem{Akkermans:2011yn}
E.~Akkermans and G.~V. Dunne, \emph{{Ramsey Fringes and Time-domain
  Multiple-Slit Interference from Vacuum}},
  \href{http://dx.doi.org/10.1103/PhysRevLett.108.030401}{\emph{Phys. Rev.
  Lett.} {\bfseries 108} (2012) 030401},
  [\href{https://arxiv.org/abs/1109.3489}{{\ttfamily 1109.3489}}].

\bibitem{Ilderton:2019ceq}
A.~Ilderton, \emph{{Coherent quantum enhancement of pair production in the null
  domain}}, \href{http://dx.doi.org/10.1103/PhysRevD.101.016006}{\emph{Phys.
  Rev. D} {\bfseries 101} (2020) 016006},
  [\href{https://arxiv.org/abs/1910.03012}{{\ttfamily 1910.03012}}].

\bibitem{tHooft:1987vrq}
G.~'t~Hooft, \emph{{Graviton Dominance in Ultrahigh-Energy Scattering}},
  \href{http://dx.doi.org/10.1016/0370-2693(87)90159-6}{\emph{Phys. Lett. B}
  {\bfseries 198} (1987) 61--63}.

\bibitem{Chadan:1996rz}
K.~Chadan, A.~Martin and J.~Stubbe, \emph{{Asymptotic properties of the
  solutions of a differential equation appearing in QCD}},
  \href{http://dx.doi.org/10.1016/0550-3213(96)00427-0}{\emph{Nucl. Phys. B}
  {\bfseries 478} (1996) 598--604},
  [\href{https://arxiv.org/abs/hep-th/9601140}{{\ttfamily hep-th/9601140}}].

\bibitem{NIST:DLMF}
``{\it NIST Digital Library of Mathematical Functions}.''
  \url{https://dlmf.nist.gov/}, Release 1.2.8 of 2026-09-15.

\bibitem{Harvey:2014qla}
C.~N. Harvey, A.~Ilderton and B.~King, \emph{{Testing numerical implementations
  of strong field electrodynamics}},
  \href{http://dx.doi.org/10.1103/PhysRevA.91.013822}{\emph{Phys. Rev. A}
  {\bfseries 91} (2015) 013822},
  [\href{https://arxiv.org/abs/1409.6187}{{\ttfamily 1409.6187}}].

\bibitem{DiPiazza:2017raw}
A.~Di~Piazza, M.~Tamburini, S.~Meuren and C.~H. Keitel, \emph{{Implementing
  nonlinear Compton scattering beyond the local constant field approximation}},
  \href{http://dx.doi.org/10.1103/PhysRevA.98.012134}{\emph{Phys. Rev. A}
  {\bfseries 98} (2018) 012134},
  [\href{https://arxiv.org/abs/1708.08276}{{\ttfamily 1708.08276}}].

\bibitem{Ilderton:2018nws}
A.~Ilderton, B.~King and D.~Seipt, \emph{{Extended locally constant field
  approximation for nonlinear Compton scattering}},
  \href{http://dx.doi.org/10.1103/PhysRevA.99.042121}{\emph{Phys. Rev. A}
  {\bfseries 99} (2019) 042121},
  [\href{https://arxiv.org/abs/1808.10339}{{\ttfamily 1808.10339}}].

\bibitem{Fedotov:2016afw}
A.~M. Fedotov, \emph{{Conjecture of perturbative QED breakdown at
  $\alpha\chi^{2/3} \gtrsim 1$}},
  \href{http://dx.doi.org/10.1088/1742-6596/826/1/012027}{\emph{J. Phys. Conf.
  Ser.} {\bfseries 826} (2017) 012027},
  [\href{https://arxiv.org/abs/1608.02261}{{\ttfamily 1608.02261}}].

\bibitem{Podszus:2018hnz}
T.~Podszus and A.~Di~Piazza, \emph{{High-energy behavior of strong-field QED in
  an intense plane wave}},
  \href{http://dx.doi.org/10.1103/PhysRevD.99.076004}{\emph{Phys. Rev. D}
  {\bfseries 99} (2019) 076004},
  [\href{https://arxiv.org/abs/1812.08673}{{\ttfamily 1812.08673}}].

\bibitem{Ilderton:2019kqp}
A.~Ilderton, \emph{{Note on the conjectured breakdown of QED perturbation
  theory in strong fields}},
  \href{http://dx.doi.org/10.1103/PhysRevD.99.085002}{\emph{Phys. Rev. D}
  {\bfseries 99} (2019) 085002},
  [\href{https://arxiv.org/abs/1901.00317}{{\ttfamily 1901.00317}}].

\bibitem{Ritus:1985vta}
V.~I. Ritus, \emph{{Quantum effects of the interaction of elementary particles
  with an intense electromagnetic field}},
  \href{http://dx.doi.org/10.1007/BF01120220}{\emph{J. Russ. Laser Res.}
  {\bfseries 6} (1985) 497--617}.

\bibitem{King:2024ffy}
B.~King and S.~Tang, \emph{{Feasibility of measuring nonanalytic QED coupling
  from pair creation in strong fields}},
  \href{http://dx.doi.org/10.1103/PhysRevA.109.032823}{\emph{Phys. Rev. A}
  {\bfseries 109} (2024) 032823},
  [\href{https://arxiv.org/abs/2401.01950}{{\ttfamily 2401.01950}}].

\bibitem{Dinu:2012tj}
V.~Dinu, T.~Heinzl and A.~Ilderton, \emph{{Infra-Red Divergences in Plane Wave
  Backgrounds}},
  \href{http://dx.doi.org/10.1103/PhysRevD.86.085037}{\emph{Phys. Rev. D}
  {\bfseries 86} (2012) 085037},
  [\href{https://arxiv.org/abs/1206.3957}{{\ttfamily 1206.3957}}].

\bibitem{Blankenbecler:1995dy}
R.~Blankenbecler and S.~D. Drell, \emph{{The Landau-Pomeranchuk-Migdal effect
  for finite targets}},
  \href{http://dx.doi.org/10.1103/PhysRevD.53.6265}{\emph{Phys. Rev. D}
  {\bfseries 53} (1996) 6265--6281}.

\bibitem{Zakharov:1996cm}
B.~G. Zakharov, \emph{{Landau-Pomeranchuk-Migdal effect for finite size
  targets}}, \href{http://dx.doi.org/10.1134/1.567248}{\emph{Pisma Zh. Eksp.
  Teor. Fiz.} {\bfseries 64} (1996) 737},
  [\href{https://arxiv.org/abs/hep-ph/9612431}{{\ttfamily hep-ph/9612431}}].

\bibitem{Baier:1997wa}
V.~N. Baier and V.~M. Katkov, \emph{{The theory of the Landau, Pomeranchuk,
  Migdal effect}},
  \href{http://dx.doi.org/10.1103/PhysRevD.57.3146}{\emph{Phys. Rev. D}
  {\bfseries 57} (1998) 3146--3162},
  [\href{https://arxiv.org/abs/hep-ph/9709214}{{\ttfamily hep-ph/9709214}}].

\bibitem{Montefiori:2026bit}
S.~Montefiori, A.~Di~Piazza, T.~Podszus, C.~H. Keitel and M.~Tamburini,
  \emph{{Intrinsic nonlocality of spin- and polarization-resolved probabilities
  in strong-field quantum electrodynamics}},
  \href{http://dx.doi.org/10.1103/cg88-lgg8}{\emph{Phys. Rev. D} {\bfseries
  114} (2026) 016030}, [\href{https://arxiv.org/abs/2603.11148}{{\ttfamily
  2603.11148}}].

\bibitem{Bonnor:1969mfs}
W.~B. Bonnor, \emph{{The gravitational field of light}},
  \href{http://dx.doi.org/10.1007/BF01645484}{\emph{Commun. Math. Phys.}
  {\bfseries 13} (1969) 163--174}.

\bibitem{Trautman:1980bj}
A.~Trautman, \emph{{A class of null solutions to Yang-Mills equations}},
  \href{http://dx.doi.org/10.1088/0305-4470/13/1/001}{\emph{J. Phys. A}
  {\bfseries 13} (1980) L1--L4}.

\bibitem{Amati:1988sa}
D.~Amati and C.~Klimcik, \emph{{Nonperturbative Computation of the Weyl Anomaly
  for a Class of Nontrivial Backgrounds}},
  \href{http://dx.doi.org/10.1016/0370-2693(89)91092-7}{\emph{Phys. Lett. B}
  {\bfseries 219} (1989) 443--447}.

\bibitem{Amati:1988tn}
D.~Amati, M.~Ciafaloni and G.~Veneziano, \emph{{Can Space-Time Be Probed Below
  the String Size?}},
  \href{http://dx.doi.org/10.1016/0370-2693(89)91366-X}{\emph{Phys. Lett. B}
  {\bfseries 216} (1989) 41--47}.

\bibitem{Horowitz:1989bv}
G.~T. Horowitz and A.~R. Steif, \emph{{Space-Time Singularities in String
  Theory}}, \href{http://dx.doi.org/10.1103/PhysRevLett.64.260}{\emph{Phys.
  Rev. Lett.} {\bfseries 64} (1990) 260}.

\bibitem{Balitsky:1995ub}
I.~Balitsky, \emph{{Operator expansion for high-energy scattering}},
  \href{http://dx.doi.org/10.1016/0550-3213(95)00638-9}{\emph{Nucl. Phys. B}
  {\bfseries 463} (1996) 99--160},
  [\href{https://arxiv.org/abs/hep-ph/9509348}{{\ttfamily hep-ph/9509348}}].

\bibitem{Iancu:2003xm}
E.~Iancu and R.~Venugopalan, \emph{{The Color glass condensate and high-energy
  scattering in QCD}}, pp.~249--3363.
\newblock World Scientific, 3, 2003.
\newblock \href{https://arxiv.org/abs/hep-ph/0303204}{{\ttfamily
  hep-ph/0303204}}.

\bibitem{Gelis:2010nm}
F.~Gelis, E.~Iancu, J.~Jalilian-Marian and R.~Venugopalan, \emph{{The Color
  Glass Condensate}},
  \href{http://dx.doi.org/10.1146/annurev.nucl.010909.083629}{\emph{Ann. Rev.
  Nucl. Part. Sci.} {\bfseries 60} (2010) 463--489},
  [\href{https://arxiv.org/abs/1002.0333}{{\ttfamily 1002.0333}}].

\bibitem{Kabat:1992tb}
D.~N. Kabat and M.~Ortiz, \emph{{Eikonal quantum gravity and Planckian
  scattering}},
  \href{http://dx.doi.org/10.1016/0550-3213(92)90627-N}{\emph{Nucl. Phys. B}
  {\bfseries 388} (1992) 570--592},
  [\href{https://arxiv.org/abs/hep-th/9203082}{{\ttfamily hep-th/9203082}}].

\bibitem{Camanho:2014apa}
X.~O. Camanho, J.~D. Edelstein, J.~Maldacena and A.~Zhiboedov, \emph{{Causality
  Constraints on Corrections to the Graviton Three-Point Coupling}},
  \href{http://dx.doi.org/10.1007/JHEP02(2016)020}{\emph{JHEP} {\bfseries 02}
  (2016) 020}, [\href{https://arxiv.org/abs/1407.5597}{{\ttfamily 1407.5597}}].

\bibitem{Kologlu:2019bco}
M.~Kologlu, P.~Kravchuk, D.~Simmons-Duffin and A.~Zhiboedov, \emph{{Shocks,
  Superconvergence, and a Stringy Equivalence Principle}},
  \href{http://dx.doi.org/10.1007/JHEP11(2020)096}{\emph{JHEP} {\bfseries 11}
  (2020) 096}, [\href{https://arxiv.org/abs/1904.05905}{{\ttfamily
  1904.05905}}].

\bibitem{DiVecchia:2023frv}
P.~Di~Vecchia, C.~Heissenberg, R.~Russo and G.~Veneziano, \emph{{The
  gravitational eikonal: From particle, string and brane collisions to
  black-hole encounters}},
  \href{http://dx.doi.org/10.1016/j.physrep.2024.06.002}{\emph{Phys. Rept.}
  {\bfseries 1083} (2024) 1--169},
  [\href{https://arxiv.org/abs/2306.16488}{{\ttfamily 2306.16488}}].

\bibitem{McLerran:1994vd}
L.~D. McLerran and R.~Venugopalan, \emph{{Green's functions in the color field
  of a large nucleus}},
  \href{http://dx.doi.org/10.1103/PhysRevD.50.2225}{\emph{Phys. Rev. D}
  {\bfseries 50} (1994) 2225--2233},
  [\href{https://arxiv.org/abs/hep-ph/9402335}{{\ttfamily hep-ph/9402335}}].

\bibitem{Ayala:1995kg}
A.~Ayala, J.~Jalilian-Marian, L.~D. McLerran and R.~Venugopalan, \emph{{The
  Gluon propagator in nonAbelian Weizsacker-Williams fields}},
  \href{http://dx.doi.org/10.1103/PhysRevD.52.2935}{\emph{Phys. Rev. D}
  {\bfseries 52} (1995) 2935--2943},
  [\href{https://arxiv.org/abs/hep-ph/9501324}{{\ttfamily hep-ph/9501324}}].

\bibitem{Gyulassy:1997vt}
M.~Gyulassy and L.~D. McLerran, \emph{{Yang-Mills radiation in
  ultrarelativistic nuclear collisions}},
  \href{http://dx.doi.org/10.1103/PhysRevC.56.2219}{\emph{Phys. Rev. C}
  {\bfseries 56} (1997) 2219--2228},
  [\href{https://arxiv.org/abs/nucl-th/9704034}{{\ttfamily nucl-th/9704034}}].

\bibitem{Balitsky:1998ya}
I.~Balitsky, \emph{{Factorization and high-energy effective action}},
  \href{http://dx.doi.org/10.1103/PhysRevD.60.014020}{\emph{Phys. Rev. D}
  {\bfseries 60} (1999) 014020},
  [\href{https://arxiv.org/abs/hep-ph/9812311}{{\ttfamily hep-ph/9812311}}].

\bibitem{Lipatov:1982vv}
L.~N. Lipatov, \emph{{Graviton Reggeization}},
  \href{http://dx.doi.org/10.1016/0370-2693(82)90156-3}{\emph{Phys. Lett. B}
  {\bfseries 116} (1982) 411--413}.

\bibitem{Raj:2023iqn}
H.~Raj and R.~Venugopalan, \emph{{Gravitational wave double copy of radiation
  from gluon shockwave collisions}},
  \href{http://dx.doi.org/10.1016/j.physletb.2024.138669}{\emph{Phys. Lett. B}
  {\bfseries 853} (2024) 138669},
  [\href{https://arxiv.org/abs/2312.03507}{{\ttfamily 2312.03507}}].

\bibitem{Raj:2023irr}
H.~Raj and R.~Venugopalan, \emph{{Universal features of $2\rightarrow N$
  scattering in QCD and gravity from shockwave collisions}},
  \href{http://dx.doi.org/10.1103/PhysRevD.109.044064}{\emph{Phys. Rev. D}
  {\bfseries 109} (2024) 044064},
  [\href{https://arxiv.org/abs/2311.03463}{{\ttfamily 2311.03463}}].

\bibitem{Raj:2024xsi}
H.~Raj and R.~Venugopalan, \emph{{QCD-gravity double-copy in the Regge regime:
  Shock wave propagators}},
  \href{http://dx.doi.org/10.1103/PhysRevD.110.056010}{\emph{Phys. Rev. D}
  {\bfseries 110} (2024) 056010},
  [\href{https://arxiv.org/abs/2406.10483}{{\ttfamily 2406.10483}}].

\bibitem{Rothstein:2024nlq}
I.~Z. Rothstein and M.~Saavedra, \emph{{A Systematic Lagrangian Formulation for
  Quantum and Classical Gravity at High Energies}},
  \href{https://arxiv.org/abs/2412.04428}{{\ttfamily 2412.04428}}.

\bibitem{Raj:2025hse}
H.~Raj and R.~Venugopalan, \emph{{QCD-Gravity double copy in Regge asymptotics:
  from $2\rightarrow n$ amplitudes to radiation in shockwave collisions}},
  \href{http://dx.doi.org/10.5506/APhysPolB.56.11-A1}{\emph{Acta Phys. Polon.
  B} {\bfseries 56} (2025) 11--A1},
  [\href{https://arxiv.org/abs/2507.21252}{{\ttfamily 2507.21252}}].

\bibitem{Stasto:2026xqw}
A.~M. Sta{\'s}to, H.~Raj and R.~Venugopalan, \emph{{Squeezed State Radiation in
  Shockwave Scattering: QCD-Gravity Double Copy}},
  \href{http://dx.doi.org/10.5506/APhysPolB.57.6-A3}{\emph{Acta Phys. Polon. B}
  {\bfseries 57} (2026) 6--A3},
  [\href{https://arxiv.org/abs/2605.03038}{{\ttfamily 2605.03038}}].

\bibitem{Blackstad:2026eoa}
I.~Blackstad, H.~Raj and R.~Venugopalan, \emph{{Tidal effects on radiation in
  gravitational shockwave collisions}},
  \href{https://arxiv.org/abs/2608.27585}{{\ttfamily 2608.27585}}.

\end{thebibliography}

\providecommand{\href}[2]{#2}\begingroup\raggedright\endgroup

\end{document}